\documentclass[sigconf]{acmart}

\newcommand{\checkinn}{\textbf{CheckINN }}
\newcommand{\checkinnnospace}{\textbf{CheckINN}}

\usepackage{listings}
\lstdefinestyle{imandra} {
	aboveskip=5pt,
	language=caml,
	keywordstyle=\color{blue},
	commentstyle=\itshape\color{purple},
	stringstyle=\color{brown},
	basicstyle=\ttfamily\footnotesize,
	breakatwhitespace=false,         
	breaklines=true,                 
	captionpos=b,       
	keepspaces=true,
	showspaces=false,                
	showstringspaces=false,
	showtabs=false,                  
	tabsize=2,
        inputencoding=utf8,
        extendedchars=true,
        literate={φ}{{$\phi$}}1
}

\usepackage{booktabs, colortbl}
\definecolor{lavender}{rgb}{0.9,0.9,0.98}

\usepackage{array}
\newcolumntype{P}[1]{>{\centering\arraybackslash}p{#1}}

\usepackage{wrapfig,multicol}

\usepackage{todonotes}
\newcommand{\knote}[1]{\todo[inline, color=blue!20]{#1}}

\AtBeginDocument{%
  \providecommand\BibTeX{{%
    \normalfont B\kern-0.5em{\scshape i\kern-0.25em b}\kern-0.8em\TeX}}}

\copyrightyear{2022}
\acmYear{2022}
\setcopyright{acmlicensed}\acmConference[PPDP 2022]{24th International Symposium on Principles and Practice of Declarative Programming}{September 20--22, 2022}{Tbilisi, Georgia}
\acmBooktitle{24th International Symposium on Principles and Practice of Declarative Programming (PPDP 2022), September 20--22, 2022, Tbilisi, Georgia}
\acmPrice{15.00}
\acmDOI{10.1145/3551357.3551372}
\acmISBN{978-1-4503-9703-2/22/09}

\begin{document}

\title{\checkinnnospace: Wide Range Neural Network Verification in Imandra}

\author{Remi Desmartin}
\orcid{}
\affiliation{%
  \institution{Heriot-Watt University}
  \streetaddress{}
  \city{Edinburgh}
  \state{}
  \country{UK}
  \postcode{}
}
\email{rhd2000@hw.ac.uk}

\author{Grant Passmore}
\affiliation{%
  \institution{Imandra Inc.}
  \streetaddress{}
  \city{Austin}
  \state{TX}
  \country{USA}}
\email{grant@imandra.ai}
\orcid{0000-0002-3240-0987}

\author{Ekaterina Komendantskaya}
\authornote{Funded by EPSRC grant AISEC (EP/T026952/1) and NCSC grant ``Neural Network Verification: in search of the missing spec.'' }
\affiliation{%
  \institution{Heriot-Watt University}
  \city{Edinburgh}
  \country{UK}
}
\email{ek19@hw.ac.uk}

\author{Matthew Daggitt}
\authornote{Funded by EPSRC grant AISEC (EP/T026952/1).}
\affiliation{%
 \institution{Heriot-Watt University}
 \streetaddress{}
 \city{Edinburgh}
 \state{}
 \country{UK}
}
\email{md2006@hw.ac.uk}

\renewcommand{\shortauthors}{Desmartin, et al.}

\begin{abstract}
Neural networks are increasingly relied upon as components of complex
safety-critical systems such as autonomous vehicles. There is high demand for
tools and methods that embed neural network verification in a larger
verification cycle. However, neural network verification is difficult
due to a wide range of verification properties of interest,
each typically only amenable to verification in specialised solvers. 
In this paper, we show how Imandra, a
functional programming language and a theorem prover originally designed for
verification, validation and simulation of financial infrastructure can offer a holistic
infrastructure for neural network verification. 
We develop a novel library \checkinn  that
formalises neural networks in Imandra, and  covers different
important facets of neural network verification.
\end{abstract}



\maketitle

\section{Motivation}
\label{sec:intro}

\begin{figure*}
	\begin{tabular}{ p{1cm} p{4.0cm}  p{3.7cm} p{2.0cm} p{3.2cm} p{2.2cm} } 
		\toprule
	& \textbf{Verification Property} & \textbf{Proof Method}  & \textbf{Type of NN} & \textbf{Matrix Representation} & \textbf{Numeric choice} \\ 
		\midrule
	Sec. \ref{sec:induct} & Structural & Induction, Imandra Waterfall & FNN, CNN  & Lists & Real, Integer \\ 
	 Sec. \ref{sec:verify} &  Reachability ($\epsilon$-ball robustness) & SAT-solver Blast & FNN, CNN & Lists &  Integer \\
       Sec. \ref{sec:acas} &  Reachability (ACAS Xu) &  Imandra Waterfall & FNN & Functions, Records & Real, Integer  \\
			\bottomrule
	\end{tabular}\caption{\footnotesize{\emph{The range of verification design decisions covered in each section of this paper.
  }}}\label{table:sections}
\end{figure*}

Machine learning algorithms have recently become a key technology
underlying complex autonomous systems such as autonomous cars, chatbots or
intelligent trading agents. \emph{Neural network} (NN) is an umbrella term for a
large family of machine-learning algorithms. Abstractly speaking, a neural
network $F$ is a function of type $\mathbb{R}^m \rightarrow \mathbb{R}^n$. We usually
understand that this function is obtained by \emph{fitting} the function's
parameters to give an optimal assignment of the available \emph{data} (given by
points in an $m$-dimensional space) to $n$ \emph{classes}. The process
of fitting such a function is usually called \emph{training} or \emph{learning},
and the optimisation algorithms used in the process are called \emph{learning
	algorithms}.

Because learning algorithms rely on incomplete and often noisy data, the
solutions they offer are difficult to verify with standard safety assurance
methods. One safety verification scenario is to prove that a neural network will
never misclassify ``important'' inputs.
This condition has several mathematical
approximations~\cite{CKDKKAE22}: e.g.\ draw an \emph{$\epsilon$-ball} around
each important data point and prove that all images within those
$\epsilon$-balls are classified correctly~\cite{HuangKWW17}. This style of
neural network analysis is often called \emph{robustness verification}, as we
prove a network robust to image change within $\epsilon$-perturbation. 
%
The verification community has proposed several algorithms
for robustness verification, a majority of which are based on either
SMT-solving~\cite{KaBaDiJuKo17Reluplex,HuangKWW17} or abstract
interpretation~\cite{SinghGPV19,GeMiDrTsChVe18,AEHW20}. The main limiting
factors for robustness verification are poor scalability to large or non-linear
neural networks, and the limited scope of $\epsilon$-ball robustness as a safety property.

Functional programming (FP) and interactive theorem provers (ITPs) have so far played only a marginal role in the
domain of neural network verification. There is a library~\cite{MariaBLFGRG22}
formalising small rational-valued neural networks in Coq and proving their structural properties. A more sizeable
formalisation called MLCert~\cite{BS19} imports neural networks from Python, treats
floating point numbers as bit vectors, and proves properties describing the
generalisation bounds for the neural networks.
An $F^*$ 
formalisation~\cite{KokkeKKAA20} uses $F^*$ reals and
refinement types for proving robustness of networks trained in Python.
Each approach had its own limitations.
For example,
MLCert does not prove neural networks' robustness, the $F^*$
formalisation only proves
robustness; neural networks in~\cite{MariaBLFGRG22} are
too small for machine learning applications that we seek to verify.

At the same time, one lesson that successful industrial provers like Imandra~\cite{PassmoreCIABKKM20,Passmore21} teach us
is that real life verification efforts require a wide range of facilities, such as (a)
user-friendly higher-order syntax, (b) ability to execute the code in order to prototype the system's behaviour and
study counterexamples, (c) proof automation for routine proofs, (d) complete techniques for bounded verification (including counterexample synthesis), and (e) facility to advance the proofs interactively when they require additional insight.
Traditionally, (a), (b), (e) can be done in ITP, and (c) and (d) in automated solvers, but often one needs all of them in the same language.
Imandra's logic is based on a pure, higher-order subset of OCaml, and
functions written in Imandra are at the same time valid OCaml code that can be
executed, or \emph{simulated}. Imandra's mode of interactive proof
development is based on a typed, higher-order lifting of the \emph{Boyer-Moore
	waterfall}~\cite{BM79} for automated induction, integrated with novel
      techniques for SMT modulo recursive functions and a first-class treatment of counterexamples.

In the present work, our main goal is to capitalise on lessons learnt by the Imandra team,
      and propose a library \checkinn~\cite{DPKD22} that provides the following four facilities that, to the best of our knowledge, no single prover has offered together before:

%


1. The choice of \textbf{properties of neural networks} that we aim to verify. Ideally, we would like to be able to prove general, higher-order properties, such as

\begin{center}\emph{$\mathcal{P^S}$: any neural network $F$ that satisfies a property $Q_1$, also satisfies a property $Q_2$.}\end{center}

\noindent Usually, proving such a property requires
induction on the structure of $F$ (as well as possibly nested induction on
parameters of $Q_1$). As such proofs rely on structural properties of $F$ captured in $Q_1$, we will call verification properties stated in this form \emph{structural properties}.
However, as this paper will show, finding such structural properties is by no means an easy task (unless the networks are small~\citeN{MariaBLFGRG22}).

This is why the verification community often resorts to proving properties like \begin{center}\emph{$\mathcal{P^R}$: for the given neural network $F$, if a property $R_1$ holds for its
inputs, verify that a property $R_2$ holds for $F$'s outputs.}\end{center} The $\epsilon$-ball robustness~\cite{SinghGPV19,GeMiDrTsChVe18,AEHW20,HuangKWW17} or ACAS Xu challenges~\cite{KaBaDiJuKo17Reluplex} are formulated in this way.  Because this kind of verification proof exploits how a property of inputs $R_1$ propagates through the given
neural network, we call verification properties stated in this form \emph{reachability properties}.

The choice of properties determines the choice of \textbf{proof methods}, which we will call respectively \emph{structural proofs} and \emph{reachability proofs}.
Crucially, Imandra can perform both structural and reachability proofs, which distinguishes it from neural network solvers like Marabou~\cite{KaBaDiJuKo17Reluplex} or ERAN~\cite{SinghGPV19,GeMiDrTsChVe18}.
Indeed Section~\ref{sec:acas} shows that Imandra uses its original proof strategies in the reachability proofs of the ACAS Xu challenge~\cite{KaBaDiJuKo17Reluplex}. However, without any further domain-specific heuristics or proved libraries of lemmas, it does
not match the performance of the domain-specific verifiers. 

2. The choice of \textbf{neural network architecture}. 
Convolutional Neural Networks (CNNs)  generalise the standard definition of ``fully connected'' neural networks (FNNs)
by introducing a range of different \emph{layer} types with different geometry.
They are widely used in computer vision as more sophisticated layer geometry allows the network to capture more general features in data.
The choice between CNNs and FNNs does not seem to play a crucial role in reachability verification, but this paper shows their
potential role in structural verification, as they open new ways of exploring the structural properties of neural networks.

CNNs are challenging for ITP formalisation, as
 they work with images and expect 2D or 3D input data, 
 and assume that different kinds of ``layers'' (convolutional, pooling, fully connected) can be composed flexibly to form a neural network, which at the level of formalisation requires a generic approach to layer definition. 
\checkinn  addresses these technical hurdles. 

\begin{lstlisting}[float=*, frame=single, caption={\footnotesize{\emph{A representative snapshot of \checkinn code for operations on matrices (as lists) and fully connected layers.}}}, label={table:fc}]
type 'a vector = 'a list
type 'a matrix = 'a vector list 
          
let dot_product (a:real matrix) (b:real matrix) =
let c = map2 ( *. ) a b in map sum c

let safe_dot_product m1 m2 = if (length m1) <> (length m2) 
		then Error "invalid dimensions" else map sum (dot_product m1 m2)

let activation f w i = (* activation function, weights, input *)
  let i' = 1.::i in (* prepend 1. for bias *)
  let z = safe_dot_product w i' in  map f z
	
let rec fc f (weights:real Matrix.matrix) (input:real Vec.vector) = match weights with
   | [] -> Ok []
   | w::ws -> lift2 cons (activation f w input) (fc f ws input)
 \end{lstlisting}


3. The choice of \textbf{matrix representation}.  No matter which neural network architecture is chosen, it still lies with the programmer to determine how
to define matrices and operations over them. Generally, functional programming languages allow many diverse approaches to representing matrices.
The standard choices are an inductive list data type~\cite{KokkeKKAA20,grant_sparse_1996}, functions from indices to elements~\cite{wood_vectors_2019}, or records~\cite{MariaBLFGRG22}.
This FP feature has already been successfully exploited in
different applications. The question we ask is whether, and how, these ideas can
be applied in neural network verification. \checkinn provides code for all three modes of matrix representation, and explores their consequences for verification. We find that some matrix representations favour reachability proofs and others structural proofs.

4. The choice between \textbf{continuous and discrete number systems}.
In theory, modelling neural networks requires working with real-valued matrices. 
Real numbers raise difficult design choices in both functional programming and theorem proving.
For constructive ITPs, the problem is in defining constructive reals~\cite{GhicaA21};
for SMT solvers --
 undecidability of real arithmetic in the presence of transcendental and special functions.
For example, Z3 uses Dual Simplex~\cite{Dutertre2006} to solve linear real arithmetic (LRA). It also supports a fragment of non-linear real arithmetic---specifically, polynomial (real-closed field) arithmetic---and solves this using a conflict resolution procedure based on cylindrical algebraic decomposition~\cite{Jovanovi2013}. However, polynomial arithmetic is not enough to cover the non-linear activation functions used in neural networks. 
Thus, solvers usually only support networks with linear activation functions (such as $relu$), and sometimes require quantisation~\cite{DuncanKSL20}.

Imandra supports real numbers via two integrated mechanisms.
For the linear case, Imandra makes use of LRA decision procedures and computation with exact rationals, including in its rewriter (in the style of Boyer-Moore \cite{boyer-moore-linear-arithmetic}). For the nonlinear case, Imandra supports reasoning with real algebraic numbers \cite{demoura-passmore-real-closed-extensions}. Moreover, as Imandra supports recursion and higher-order functions, non-polynomial real functions may be defined and reasoned about by defining recursive functions which approximate them via, e.g., Cauchy sequences.

\checkinn capitalises on Imandra's real number facilities and this paper makes a point of studying where, and how, transitioning between real-valued and quantised matrices makes a difference for neural network verification.



 The paper is structured as follows.
Section~\ref{sec:background} gives necessary background on neural networks and Imandra syntax.
Section~\ref{sec:CNN} introduces an implementation of CNNs in Imandra. 	Section \ref{sec:induct} presents the first verification task -- a proof of a structural property (neural network monotonicity) for FNN, and at the same time illustrates Imandra's waterfall method. It then considers a more difficult scenario of formulating and proving structural properties of CNNs.
Section~\ref{sec:verify} uses Blast, Imandra's SAT solver on the matrix-as-list representation, to prove $\epsilon$-ball robustness for small networks. But Blast only works with integers and does not scale well to big neural networks. Finally, Section~\ref{sec:acas} uses Imandra's native automation to solve the ACAS Xu challenge\cite{KaBaDiJuKo17Reluplex} with integer and real values,  but this comes at the price of working with matrix representations via maps and records.
Section~\ref{sec:concl} concludes the paper. 



The main contribution of this paper is to demonstrate the power of the ``wide range'' NN verification methodology in \checkinnnospace.
Fig.\ref{table:sections} summarises the combination of methods explored across the sections. 
This main contribution builds upon several smaller original contributions, e.g. the formalisation of Section~\ref{sec:CNN} is the first ITP implementation of CNNs we are aware of, the method of structural verification of CNN proposed in Section \ref{sec:induct} is original; Sections~\ref{sec:verify} and \ref{sec:acas} are the first successful attempts to automatically prove reachability properties for common benchmarks in any ITP.

\section{Background}\label{sec:background}

In this section, we introduce fully connected networks. 
Typically one works with neural networks formed by composition of \emph{layers}. So, we start with defining layers first. 
Given two matrices $w$ and $b$ that are called a \emph{weight} and a \emph{bias},  
a \emph{layer}  $L$ is a function defined as:
\begin{equation}
	L(x) = a(x \cdot w + b)
\end{equation}
where the operator $\cdot $ denotes the dot product between the input vector $x$ and each row of $w$, 
 and $a: \mathbb{R} \rightarrow \mathbb{R}$ is the activation function applied pointwise to the elements of the vector obtained by computing $(x \cdot w + b)$. 

By denoting $a_k, w_k, b_k$ --- the activation function, weight and bias of the $k$th layer respectively, a \emph{fully-connected network (FNN)} $F$ with $l$ layers is traditionally defined as:
\begin{equation}
  F(x) = L_l(L_{l-1}(\ldots L_1(x) \ldots))
\end{equation}


 In Imandra, we aim to define NNs as functions that compose layers:

\begin{lstlisting}[frame=single, language=caml, label={lst:model}, caption={\footnotesize{\emph{Desirable Syntax for NN using monadic bind (cf. Appendix~\ref{ap:monad}).}}}]
let cnn input =
      layer_0 input >>= layer_1 >>= layer_2 >>= layer_3
\end{lstlisting}






 To implement this in FP, we have three key choices:
\begin{enumerate}
\item to represent matrices as lists of lists (and take advantage of the inductive data type \lstinline{List}),
\item define matrices as functions from indices to matrix elements,
\item or take advantage of record types, and define matrices as records with maps.  
\end{enumerate}
In the accompanying note~\cite{DPK22} we focus specifically on the technical consequences of taking each of these choices in Imandra. Here, we will build up our matrix representations gradually, explaining the consequences of various choices as we go.

We start with lists (cf. Listing~\ref{table:fc}).
%
Most list manipulation functions of the OCaml \lstinline{List} module are available in Imandra, which opens the way for library re-use.
For example, the definition of a dot product uses  \lstinline|map2| (a straightforward generalisation of list \lstinline{map} to matrices)
and the \lstinline|sum| function, which in turn applies \lstinline|List.fold_left| and  \lstinline|+| to vectors.
A fully connected layer is then defined as a function \lstinline{fc} which takes
as parameters an activation function, a 2-dimensional matrix of
layer's weights and an input vector. 
Note that each row of the weights matrix
represents the weights for one of the layer's nodes. The bias for each node is
the first value of the weights vector, and $1$ is prepended to the input vector
when computing the linear combination of weights and input to account for that.



It is now easy to see that our desired approach to composing layers given in Listing~\ref{lst:model} works as stated. We may define the layers using the syntax:
\lstinline{let layer_i = fc a weights}, where \lstinline{i} stands for \lstinline{0,1,2,3}, and \lstinline{a} stands for any chosen activation function. 

Although natural, this formalisation of layers and networks suffers from two problems.  Firstly, it lacks the matrix dimension checks that were readily provided  via refinement types in~\cite{KokkeKKAA20}. This is because Imandra is based on a computational fragment of HOL, and has no refinement or dependent types. To mitigate this, the library we present performs explicit dimension checking via a {\tt result} monad (indeed the code in Table~\ref{table:fc} gives a good idea of dimension error tracking in this part of \checkinnnospace).
Secondly, the matrix definition via the list data types makes unrolling-based~\cite{PassmoreCIABKKM20} proofs of robustness inefficient, as even accessing matrix elements typically involves unfolding several layers of recursion.
In Section~\ref{sec:acas} we will present a more efficient approach that defines matrices as functions, and alleviates both of these problems.
However, we proceed with this simpler data type definition of matrices for the time being. 


\begin{lstlisting}[float=*, frame=single, caption={\footnotesize{\emph{A representative snapshot of \checkinn code for CNN.}}}, language=caml, label={lst:convolution}]
let rec convolution_row' input filter (row, col) =
    let (row', col') = Matrix.dimensions filter in
    if col < 0 then Ok [] else
    let sub_m = Matrix.sub_matrix input (row, col) (row', col') in
    let dot_p = Res.bind sub_m (fun x -> Matrix.dot_product x filter) in
    let head = convolution_row' input filter (row, col - 1) in (* col decreases to let imandra prove termination *)
    Res.bind2 head dot_p (fun x y -> Ok (x @ [y]))

let convolution_row input filter row =
    let (i_rows, i_cols) = Matrix.dimensions input in
    let (f_rows, f_cols) = Matrix.dimensions filter in
    if i_rows < f_rows then Error "convolution_row: filter's height is greater than input's" else
    if i_cols < f_cols then Error "convolution_row: filter's width is greater than input's" else
    let col = (i_cols - f_cols) in 
    convolution_row' input filter (row, col)
        
let convolution (input: real Matrix.t) (filter: real Matrix.t) = 
    if not (Matrix.is_valid input) || not (Matrix.is_valid filter) then Error "convolution: invalid matrix" else
    let (i_rows, _) = Matrix.dimensions input in
    let (f_rows, _) = Matrix.dimensions filter in
    if i_rows < f_rows then Error "convolution: filter's size is greater than input's " else
    let acc_fun (i, xs) _ =                               
        if f_rows + i > i_rows then (i + 1, xs) else       
        let x = convolution_row input filter i in
        (i + 1, Res.bind2 x xs (fun x xs -> Ok (xs@[x]))) in
    let (_, res) = List.fold_left acc_fun (0, Ok []) input in 
    res
\end{lstlisting}

\section{CNNs  in Imandra}\label{sec:CNN}

\begin{figure}
	\centering
	\includegraphics[width=0.45\textwidth]{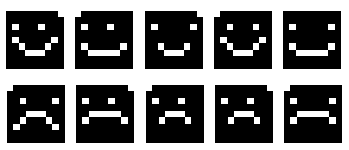}
	\caption[Dataset examples]{\footnotesize{\emph{Sample from the dataset used as a running example. The images in the top row are labelled as "Happy" and those in the bottom as "Sad".}}}
	\label{fig:dataset_sample}
      \end{figure}

\begin{figure}
	\vspace{-0.2em}
	\centering
	\includegraphics[width=0.45\textwidth]{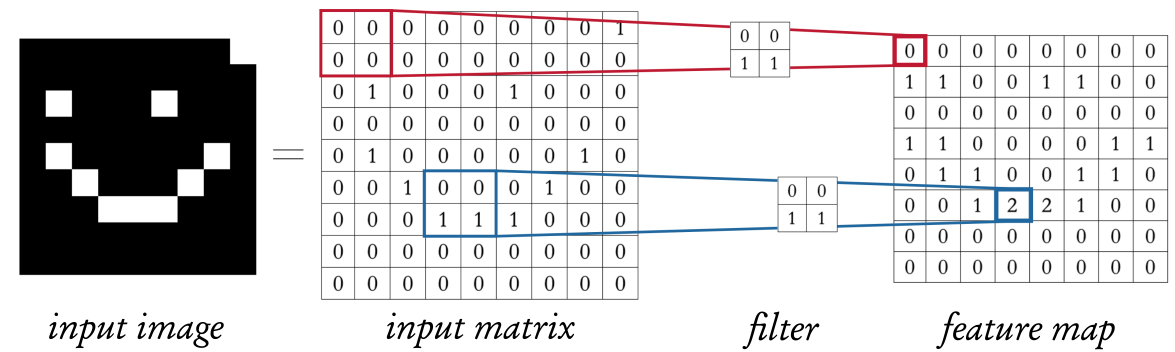}
	\caption[Convolution]{\footnotesize{\emph{A feature map resulting from a convolution operation, given an image. The filter shows a horizontal line pattern.
				The bottom area of the input image matches the filter better than the top one, resulting in higher values in the feature map.}}}
	\label{fig:convolution}
\end{figure}

\begin{figure}
	\centering
	\includegraphics[width=0.25\textwidth]{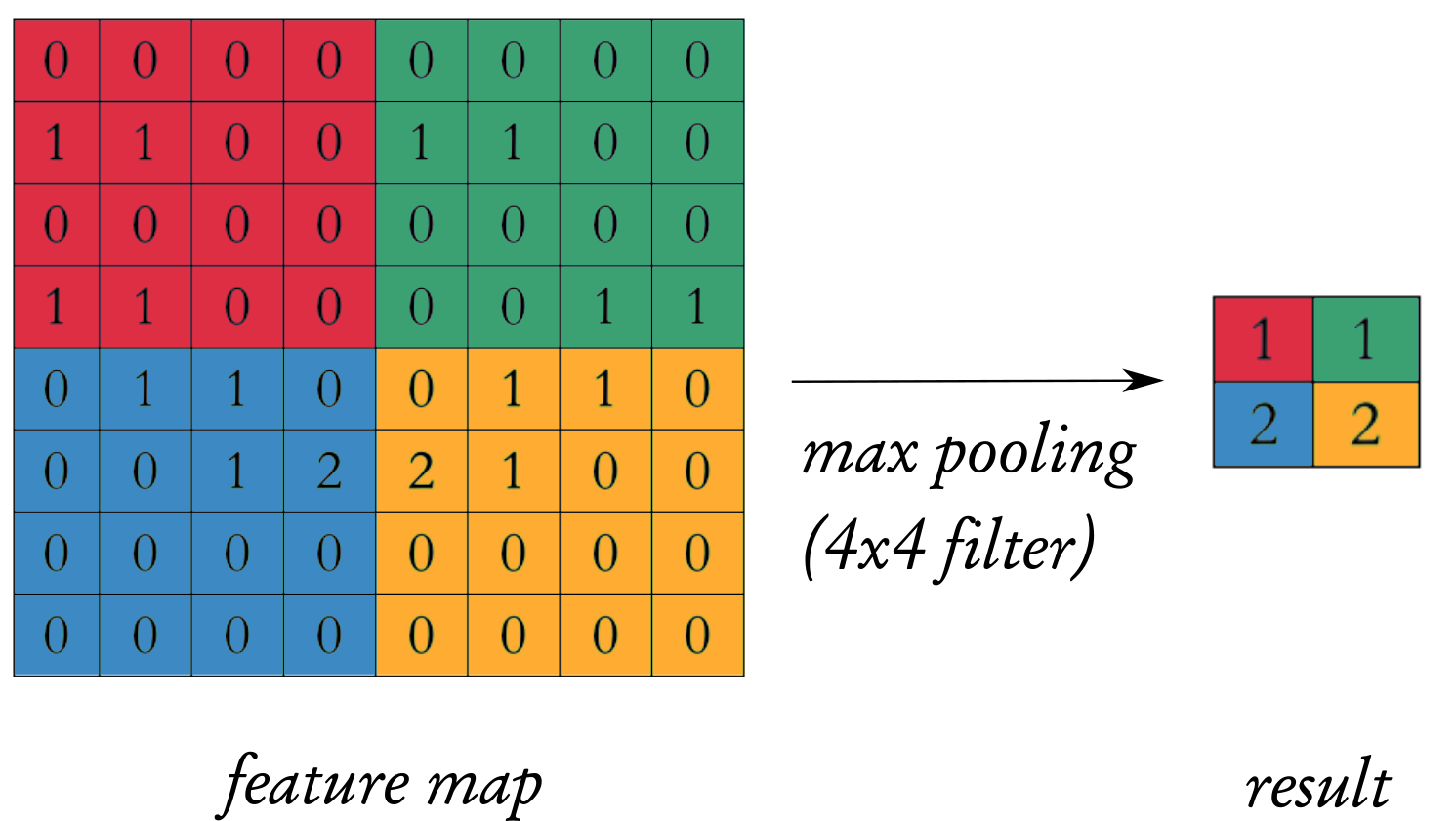}
	\caption[Max Pooling]{\emph{\footnotesize{Max pooling operation with a $4\times4$ filter. Each coloured zone is a region where the filter is applied.}}}
	\label{fig:max_pooling}
      \end{figure}


  \begin{figure*}[t]
	\centering
	\includegraphics[width=0.7\textwidth]{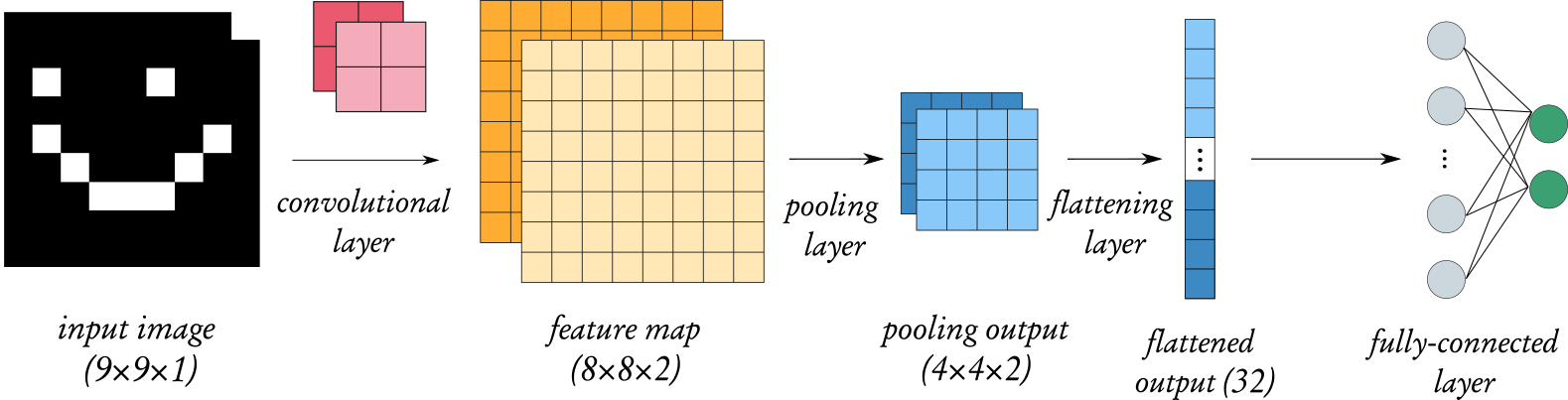}
	\caption[Convolutional Network Example]{\footnotesize{\emph{Representation of a CNN's layers and intermediate outputs with their dimensions.}}}
	\label{fig:cnn_layers}
\end{figure*}


We now introduce the CNN part of \checkinnnospace. Unlike their fully-connected
counterparts, CNNs are designed to make use of spatial information that may be
present in data. To illustrate this, consider the following artificial data set created
to serve as a running example for this paper. It consists of 144 unique
images of dimension $9\times9\times1$ (see Fig.~\ref{fig:dataset_sample}), in
which images are classified as happy or sad faces. This toy data set makes it easier for us to expose the main ideas behind CNN verification.
In later sections, we will be using ACAS Xu data set and networks~\cite{KaBaDiJuKo17Reluplex} as well.

The best way to recognise a
smile or a frown is by considering the spatial configuration of pixels around
the mouth. If these pictures were flattened into vectors, the spatial
information would be lost. 
In order to analyse 2D data, layers of different kinds (convolutional,
pooling and fully connected) operate over the
submatrices of the matrix that represents a given data point, as illustrated in
Fig.~\ref{fig:cnn_layers}. We first describe how the Imandra library defines
each of these layers.

%
\subsection{Convolutional Layer} \label{sct:backgrond_convolutional}

Convolutional layer weights are given by several  multi\-dimensional matrices, called \emph{filters}.
Each filter captures some distinct ``feature''. For example, given three filters, each can detect respectively diagonal,  vertical and  horizontal lines present in an image.  
The layer output is the result of convolution operations between the input image and each of its filters. For each filter, the layer outputs one 2-dimensional array called a \textit{feature map}. 	Fig.~\ref{fig:convolution} shows a convolution operation between an image and a filter.


\begin{definition}[Convolution Operation]\label{def:CN}
	Let $J$ be a 2-dimensional matrix of size $h_j \times w_j$, and let $K$ be a 2-dimensional square \emph{filter} of size $k \times k$,
	then the \textit{feature map} $M$ is the result of the \emph{convolution operation} between $J$ and $K$, defined as follows.
	$M$'s dimensions are $(h_j - k + 1 )\times (w_j - k + 1)$, and the value of its elements at the intersection of the $i^{th}$ row and $n^{th}$ column is determined by the equation:
	%
	$M_{i,n} = \sum_{s=1}^{k} \sum_{p=1}^{k} K_{s, p}J_{(i + s), (n + p)}$.
\end{definition}

\noindent To make it more amenable to formalisation in Imandra, we will slightly rewrite this definition.
By using $X[i_1,i_s; n_1,n_t]$ to denote the submatrix of a matrix $X$ formed by the intersection of the rows $i_1$ to $i_s$ and columns $n_1$ to $n_t$, we use:
%
$$M_{i,n} = K \cdot J[i,i+k;n,n+k].$$


Because filters are intended to represent features, i.e.\ patterns characteristic of a class, 
a convolution operation between filters and an input matrix can be seen as checking which part of the input matches the feature present in the filter; hence the name ``feature map'' for its result.  
%
%
We formalise the convolutional layer in Listing~\ref{lst:convolution}.
The function 
\lstinline{convolution} implements a convolution operation between a matrix and a filter; \lstinline{fold_left} iterates the operation.

So far, we assumed that the input matrix only has one colour channel, but images usually have three. 
To apply a convolution operation to input with multiple channels, the filters must have the same number of channels. \checkinn~ handles such cases.

\subsection{Pooling Layer}


      Pooling layers come after convolutional layers; their input is the
feature maps from the previous layer, and their output is a set of 2-dimensional
matrices that reduce feature maps in size.
Two main types of pooling operations are used in CNNs:
max pooling and average pooling. By abuse of terminology, the literature
also refers to filters in the pooling layer (cf. Fig~\ref{fig:max_pooling}),
but in fact ``filters'' here simply define the submatrix size. Our formal definition clarifies this point.

\begin{definition}[Max Pooling Operation]\label{def:PL}
	Given a 2-dimensional input matrix $J$, and the \emph{filter of size k}, the \emph{max pooling operation}
	is a function that returns a matrix $M$, whose elements are defined by
	$M_{i, n} = max(J[i,i+k;n,n+k])$.
\end{definition}

The \checkinn implementation of pooling layers closely mimics
the style of formalisation of the convolutional layer, except for using
the $max$ operation instead of the dot product. Average pooling layers work the
same way, but instead of a $max$ function, they use an averaging function.




\subsection{Assembling All Layers}

CNNs alternate between convolutional layers and pooling layers. This allows them to achieve a ``greater level of abstraction''
in feature detection. A typical CNN architecture usually chains several convolutional layers and pooling layers.
The \emph{flattening layer}  flattens the 3D representation into a vector, and several fully connected layers complete the network.

We can now assemble the network shown in Fig.~\ref{fig:cnn_layers}, by using the syntax of Listing~\ref{lst:model}. To do this, we need to import a trained neural network from Python.
\checkinn uses Keras to train our networks, and it contains a Python script to convert a CNN saved in Keras~\cite{chollet2015keras} format into an Imandra module containing each layer's weights. The layers must then be instantiated with layer functions. 
The \lstinline{Layers} module encapsulates the individual layer modules to expose higher-order functions that instantiate layer functions. The convolution and max pooling layers are implemented in their respective modules for a single filter but \lstinline{Layers} can hold several filters; all filters are then \emph{flattened} together.
These functions are  partially applied to a multi-dimensional array of weights, 
to create layer functions that can be chained to form a network: 

\begin{lstlisting}[label={lst:model2}]
	let layer_0 = Layer.convolution Layer0.filters
	let layer_1 = Layer.max_pool (2, 2)
	let layer_2 = Layer.flatten
	let layer_3 = Layer.fc (fun x -> x) Layer3.weights
	let model input = layer_0 input >>= layer_1 >>= layer_2 >>= layer_3
\end{lstlisting}


As these layer functions are implemented in a generic way, an arbitrary number of layers of any type can be chained together as long as the dimensions of each layer's output match those of the next expected input. (The dimensions are checked dynamically, and we will see errors at run time if the dimensions do not match.)
For instance, an FNN can be created by chaining only fully connected layers.
Note that the dimensions of layer inputs and outputs are not specified in the user interface to the library, they are deduced from the layer dimensions. 



%
%

\section{Structural Properties}\label{sec:induct}


When proving structural properties of neural networks, we are interested in showing how a certain feature present in
a network's architecture influences its behaviour as a function. As a consequence, such proofs quantify over all neural networks with said
architectural features; and usually require induction on the network's structure. This style of proof matches best with Imandra's original design,
as a higher-order inductive theorem prover. 

We start with a simple example of a monotonicity property for a FNN 
and use it to also illustrate
 Imandra's \emph{Waterfall} proof method. We then investigate structural properties of CNNs that may be useful in verification.

\subsection{Monotonicity and Inductive Proofs}\label{subsec:mono}

 There has been some
interest in monotone networks in the literature~\cite{JS98,WehenkelL19}. We will emulate a monotonicity property as follows: any fully connected
network with positive weights is \emph{monotone}, in the sense that, given
increasing positive inputs, its outputs will also increase. 

For the sake of this section, we somewhat simplify the code for FNNs. Taking lists of real numbers for input \lstinline{i},
 2D and 3D matrices (as lists) for the weights (\lstinline{ws}) and biases (\lstinline{bs}), we define:

\begin{lstlisting}
let rec layer ws bs i = match (ws, bs) with
   | (_, []) | ([], _) -> []
   | (w::ws, b::bs) -> (perceptron' w b i) :: (layer ws bs i)
	
let rec network ws bs i = match (ws, bs) with
   | (_, []) | ([], _) -> i
   | (w::ws, b::bs) -> network ws bs (layer w b i)   
\end{lstlisting}



The monotonicity theorem is then stated simply as: 

\begin{lstlisting}
theorem network_monotonicity ws bs i i' =
positiv_3d ws && positiv_2d bs && positiv i && gte i' i
==> gte (network ws bs i') (network ws bs i)
      \end{lstlisting}
\noindent where the positivity conditions at the top
are for vectors, matrices and 3D matrices respectively; and \lstinline{gte} stands
for ``greater or equal'' applied pointwise to list elements. Note that the theorem above quantifies over FNNs of any size.


Imandra's proofs are based on the \emph{Boyer-Moore
  waterfall}~\cite{BM79} strategy, which automates induction, by generating plausible induction principles, and also applying several heuristics to discharge intermediate goals.
The waterfall (\lstinline{[@@auto]}) proceeds in five steps: 
\begin{enumerate}
  \item \textbf{simplification} makes use of all enabled rewrite and forward-chaining rules, decision procedures for algebraic data types and arithmetic, and case-splits,
  \item \textbf{definition unrolling} searches for counterexamples up to a certain unrolling depth (not unlike say in QuickCheck~\cite{CH00}),
  \item \textbf{destructor elimination} transforms all expressions of inductive types from a destructor form (e.g \lstinline{a = List.hd x} and \lstinline{b = List.tl x}) into a constructor form (e.g. \lstinline{x = a::b}),
  \item \textbf{fertilisation} performs rewriting on terms in the goal using equivalent terms defined in the lemma assumptions,
  \item \textbf{generalisation} generalises the given conjecture.
\end{enumerate}
Then Imandra generates an induction scheme for the generalised goal and restarts the search for a proof from item 1.
Although much of this process is automated, Imandra switches to an interactive mode when the proof search fails and suggests missing lemmas. 

    Let us see Imandra's waterfall in action (see also Appendix~\ref{app:code} for full user dialogue).
    The proof of monotonicity does not succeed immediately by \lstinline{[@@auto]}. On the first attempt, Imandra realises it has to use induction, and finds six possible ways to proceed by induction. It however manages to simplify the six to two, and finds a clear winner among the two, with induction on the structure of \lstinline{ws} and \lstinline{bs}. 
To finish the proof, we prove two
lemmas (derived by analysing Imandra's ``simplification checkpoints'' for the goal): 

\begin{lstlisting}
	lemma positive_push_2d bs ws i =
	positive bs  && positive_2d ws && positive i
	==> positive (layer ws bs i) [@@auto] [@@rw]
	
	lemma layer_monotonicity ws bs i i' =
	positive_2d ws && positive bs && positive i && gte i' i
	==> gte (layer ws bs i') (layer ws bs i) [@@auto][@@rw]
\end{lstlisting}

\noindent The first lemma shows that positivity is inherited from layer inputs
to layer outputs; and the second asserts the monotonicity property for
individual layers. Then Imandra completes the inductive proof by
\lstinline{[@@auto]}. 
Automation of inductive proofs is a strength of Imandra. Similar Coq proofs
in~\cite{MariaBLFGRG22} required more tactic guidance.

\subsection{Structural Properties of CNNs}\label{subsec:structural_property}


It may seem that general structural properties like monotonicity are not very useful for practical verification tasks, especially in CNN verification.
In this section, we present an example that shows a structural approach to CNN verification and exposes how general mathematical properties like monotonicity can play a role in the process. 
We continue to work with the same toy image data set and 
the CNN $F$ (of Fig.~\ref{fig:cnn_layers}). But we highlight the general pattern in the approach that can be extrapolated to other CNNs.



\begin{figure}
	\centering
	\includegraphics[width=0.12\textwidth]{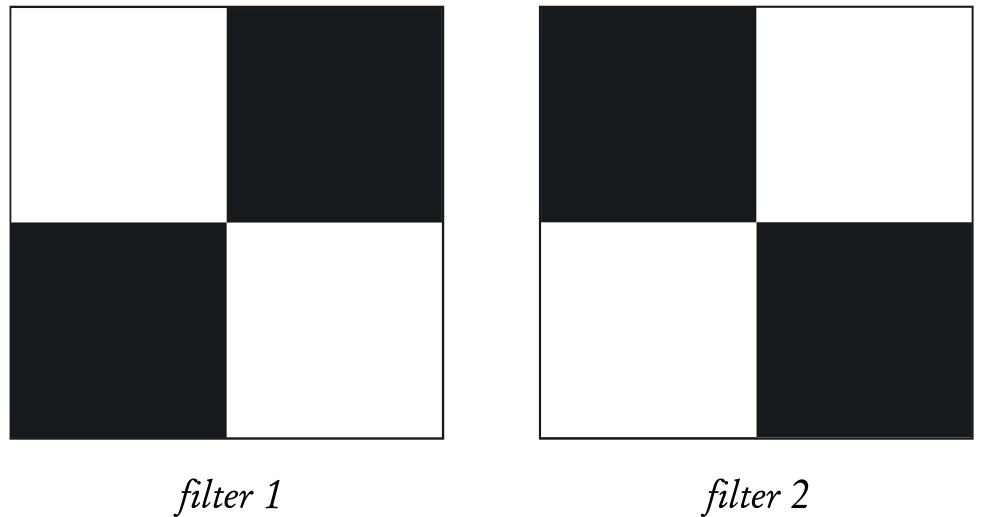}
        	\hspace{1.2cm}
                \includegraphics[width=0.12\textwidth]{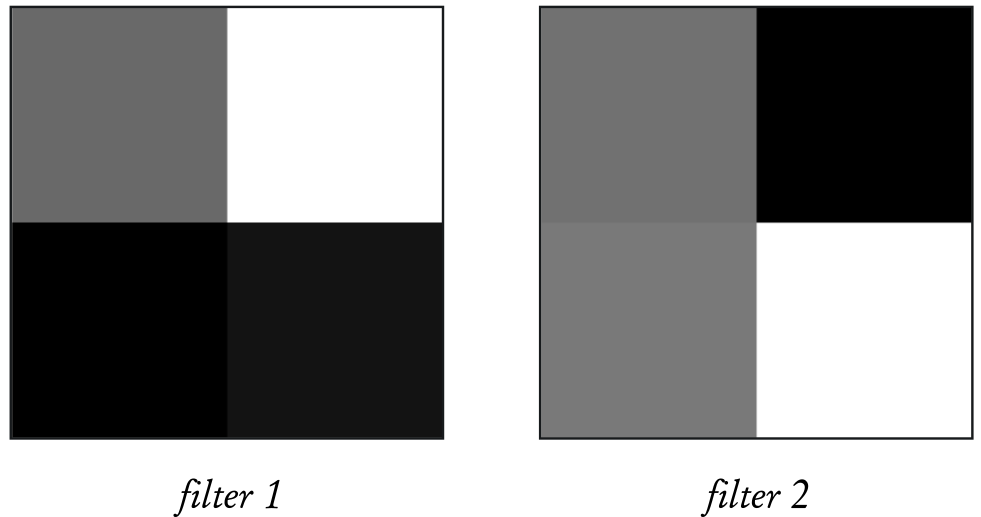}
                
                \caption[Imposed filters visualisation]{\footnotesize{\emph{Heatmaps of $2\times2\times1$ filters: human-imposed on the left; learnt by the CNN on the right.
                    }}}
	\label{fig:diagonal_filters}
      \end{figure}

      \begin{figure}
	\centering
	\includegraphics[width=0.4\textwidth]{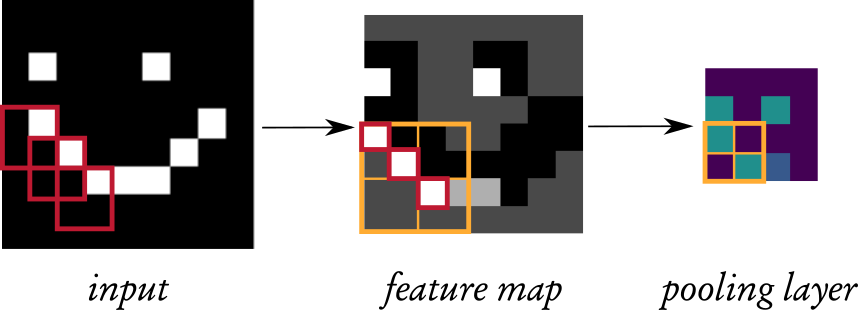}
	\caption[Designing pooling layer properties]{\emph{\footnotesize{A filter propagated through CNN layers.}}}
	\label{fig:properties_design}
      \end{figure}




\textbf{1. Filter Adequacy.}
We assume that filters bear some structural meaning. 
For example, a set of CNN filters that would agree with the human definition of a smile could give rise to \emph{heatmaps}
that show diagonal lines, as on the left of Fig.~\ref{fig:diagonal_filters}. These diagonals are then detected in the later layers of the CNNs, as Fig.~\ref{fig:properties_design} shows.\footnote{
Recall that filters are real matrices;  heatmaps are visualisations of such matrices, where intensity of colours corresponds to values of matrix elements.
In particular,  in Fig.~\ref{fig:diagonal_filters}, black stands for $0$ and white for $1$.}

In reality, the situation is a bit more complicated. When we train a 100\% accurate CNN on this data set and examine heatmaps of the filters it learnt, we notice that in fact it learnt the filters shown on the right of Fig.~\ref{fig:diagonal_filters}.
It is not immediately clear how to interpret what the CNN thinks a smile is. However, digging deeper into the layers, we found that both manually constructed and learnt filters give rise to a well-defined pattern in the pooling layer, so it seems that analysis of the pooling layer is crucial. We thus want to define a filter to be \emph{adequate for recognition of a smile} if it gives rise to
a well-defined \emph{pattern} in the pooling layer, given a smiling face as an input. The pattern is given by two small diagonals in the bottom corners of the matrix:

\begin{lstlisting}
let has_pattern (m: (('a Matrix.t) Vec.t)): (bool, 'b) result = Ok (max_bottom_left_corner (Vec.nth 0 m) &&
        max_bottom_right_corner (Vec.nth 1 m))
      \end{lstlisting}


\textbf{2. Definition of Verification Property.}
Next, we assume that a definition of a smiling face is a specification that can be written by a human.
Such a specification usually captures idealised structure and abstracts away from any exceptions.
For example, for the given data set, \checkinn
 defines a happy face as the one that smiles, 
and a smile as \emph{a shape with a left diagonal and a right diagonal on either side of the
mouth region, connected by a horizontal line}.

The need for neural networks arises when we want to implement systems that deal with noise and exceptions.
For example, the picture may be distorted or taken from a wrong angle.
In such cases, an idealised rule would fail, whereas a neural network may still succeed.
In our verification scenario, we want to ensure some form of soundness, i.e. prove that all cases that fall under the human specification of a happy face
are classified as happy:  \emph{given a CNN F with adequate filters and a well-tuned fully-connected layer,
        any image that satisfies the specification of a happy face will always be classified by F as happy}.

     

    The theorem misses the definition of a well-tuned fully-connected layer. From the engineering point of view, the weights of that layer must be tuned to higher values
    exactly where the ``hot'' pattern in the pooling layer of an adequate filter is expected. And this is the point that requires a more general mathematical reasoning. 

    \textbf{3. Extreme Values Lemma.}
    To understand the problem recall the role of the fully connected layer in a CNN.
    Let $x$ and $y$ denote the two neurons of the output layer $out$, standing for classes ``Happy'' and ``Sad''.
    Each of these neurons represents the score for a class for a given input image.
    The weights associated with $x$ and $y$ are respectively denoted by $w^x = (w^x_0, w^x_1,..., w^x_n)$ and $w^y = (w^y_0, w^y_1,..., w^y_n)$. 
    After a certain pattern in a (happy) image is detected in the pooling layer, the pooling layer is flattened into a vector of weights which are used to compute the vector that the units $x$ and $y$ receive. 
    So, ultimately, it is now up to neurons $x$ and $y$ to classify the pooling layer pattern into one of the two classes.

    Let $a$ denote the input vector of the layer $out$ and $a^*$ -- the mean value of $a$.
    Without loss of generality, let us ignore the activation function of the layer $out$, and just concentrate on its dot products. For unit $x$, it calculates
$	(a_1 w^x_1 + ... + a_n  w^x_n)$ and for unit $y$  it calculates  $(a_1  w^y_1 + ... + a_n w^y_n) $.
It then decides on the class of the CNN input based on checking

$$	(a_1  w^x_1 + ... + a_n w^x_n) > (a_1  w^y_1 + ... + a_n w^y_n) $$

We are almost led to believe that we simply need some monotonicity property that shows that $w^x > w^y$. However, this property would not apply to CNNs we find in practice. In reality, each of the weight vectors $w^x$ and $w^y$ has higher values for certain indices (and lower for others) depending on which features are characteristic of which class.
Intuitively, if we know that a smile means high values in the bottom corners of the pooling layer, then it is specifically for these regions that $w^x$ will have higher values than $w^y$. So, the general property that we need should describe how well the fully-connected layer is tuned to these ``extreme values''. 

    Let $a_{max}$ and $a_{min}$ be the max and min values in $a$.   
We define extreme values of $a$ as 
\begin{equation}
	a_{ex} = \{ a_i : a_i > a^* + \frac{a_{max} - a^*}{2} \}
      \end{equation}

      We say a vector $a$ \emph{has a distinct pattern} if all values of $a$ are either extreme or below $a^*$. 



\begin{lemma}[Extreme Values]\label{lem:extreme}
Let $a$ be a vector $a_1, \ldots , a_n$ with distinct pattern and extreme values $a_1, \ldots , a_m$. If for  $w^x$ and $w^y$, we have 
$$
	w^x_1 > w^y_1 \land ... \land w^x_m > w^y_m,
$$

then 
$$
	(a_1  w^x_1 + ... + a_n  w^x_n) > (a_1  w^y_1 + ... + a_n w^y_n) .
$$

\end{lemma}

Just as in the case of monotonicity, we note that the lemma is stated in full generality, and does not depend on a specific network architecture or a data set. 
Without further constraining the vectors $a$, $w^x$ and $w^y$, the lemma does not hold. Defining necessary restrictions on these vectors ultimately gives
possible definitions of the ``well-tuned fully-connected layer''.
Indeed, Imandra's facility for counter-example generation may serve as an aide in formulating the new conditions. 

In particular, the lemma holds for the following two special cases (see the proofs in Appendix~\ref{ap:extreme} or in \checkinn):
\begin{itemize}
\item[\textbf{R1. }] $a$ is a binary vector, and $a_{min} \neq a^*$, $a_{max} \neq a^*$; 

\item[\textbf{R2. }]  $a$ has positive values, $w^x$ and $w^y$ are binary vectors, and $m \geq \frac{n}{1.5 + \frac{a_{max}}{2a^*}}$.
  
\end{itemize}

\begin{lstlisting}[float=*, frame=single, caption={\footnotesize{\emph{Imandra's proof of the bounded version of Extreme Values Lemma, Case \textbf{R2}}}}, label={lst:bound}]
lemma extreme_value_lemma_r2_len_8 x1 x2 x3 x4 x5 x6 x7 x8 y1 y2 y3 y4 y5 y6 y7 y8 a1 a2 a3 a4 a5 a6 a7 a8 =
  let w_x, w_y, a = [x1; x2; x3; x4; x5; x6; x7; x8],
                    [y1; y2; y3; y4; y5; y6; y7; y8],
                    [a1; a2; a3; a4; a5; a6; a7; a8]
  in
  extreme_value_precondition_r2 w_x w_y a ==> extreme_value_postcondition w_x w_y a
[@@unroll 200]
\end{lstlisting}

Manual proofs of these two cases are short (see Appendix~\ref{ap:extreme}), but assume some facts about the relations between $a^*$, $a_{max}$, $a_{mean}$, $a_{ex}$, which would be laborious to formalise. Our methodological interest here is to show that Imandra can ease one's verification tasks, rather than complicate them. For \textbf{R1}, we can formalise the lemma in a way that will guide Imandra's inductive proof in the right direction. In particular, we can incorporate our knowledge of extreme values into definition of the dot product:

\begin{lstlisting}

type value =
  | Extreme of (real * real * real)
  | Normal of (real * real * real)

type vecs = value list

let rec dot_products vs =
  let open Real in
  match vs with
  | [] -> 0., 0.
  | Extreme (x_i, y_i, a_i) :: vs
  | Normal (x_i, y_i, a_i) :: vs
    ->
    let (p1, p2) = dot_products vs in
    (x_i *. a_i +. p1,
     y_i *. a_i +. p2)

let is_valid_r1 vecs =
  let rec aux seen_ex vecs =
  let open Real in
  match vecs with
  | [] -> seen_ex
  | Extreme (x_i, y_i, a_i) :: vs ->
    a_i = 1.0
    && x_i > y_i
    && aux true vs
  | Normal (x_i, y_i, a_i) :: vs ->
    a_i = 0.0
    && aux seen_ex vs in
  aux false vecs

lemma main vs =  is_valid_r1 vs   ==>
  let (p1,p2) = dot_products vs in  p1 >. p2
  [@@auto]
\end{lstlisting}

\noindent Imandra's proof of the main lemma (see Appendix~\ref{ap:extreme}) is long but elegant: Imandra is able to automatically discover  an inductive proof scheme that relates the vector size and extreme values!

For \textbf{R2},  there is no easy way to play a similar trick, and we need to formalise all lemma preconditions in full generality. They are fairly straightforward albeit lengthy, and can be found in the appendix. However, this time Imandra cannot complete the proof automatically. Once again, we will try to avoid burdensome auxiliary lemmas, and instead showcase yet another useful Imandra tactic: \lstinline{[@@unroll]}.

We already mentioned that unrolling in Imandra plays a role of a counterexample finder. But, since it is based on the idea of symbolic bounded model checking modulo ground decision procedures,
there is another way it can be used in proofs: we can prove results over bounded structures, even if these structures contain, e.g., unbounded reals or integers.
The key point is that with a fixed explicitly given list structure, all recursive functions can be completely unrolled and eliminated by Imandra, and what is left is a ground SMT problem which is amenable to decision procedures.
Listing~\ref{lst:bound} shows the full interaction with Imandra for proving \textbf{R2} for vectors of dimension $8$, which is the maximal dimension we were able to verify with a 300-second timeout. 
This form of bounded verification is very useful for analyzing concrete conjectures, and may suffice for many verification scenarios in which the architecture of networks is known in advance.


From the methodological point of view, conditions like \textbf{R1} and \textbf{R2} give us a way to construct CNNs that can in principle be proved sound. For example, to obtain CNNs that satisfy \textbf{R1}, we would need to apply a binary threshold function on activations of the pooling layer.
To obtain \textbf{R2}, we would need to use algorithms that binarise the weights when training.
Having these conditions, assembling the other components of the soundness theorem is trivial.

It is out of the scope of this paper to seek more liberal restrictions to the Extreme Values Lemma;
however, this would be the future line of work for any scalable project on structural verification of CNNs that follows the described verification scenario.
The restrictions above suggest that the key to extending the lemma's applicability
is to find more sophisticated formulae that characterise the relationship between the magnitude and the number of extreme values. 

\section{Reachability and Symbolic Execution} \label{sec:verify}

    \begin{figure*}
	\begin{tabular}{p{5.5cm} p{7.5cm}} 
		\toprule
		\textbf{Property} & \textbf{Formal definition} \\ 
		\midrule
		Classification robustness (CR) & 	$\forall X: ||X-\hat{X}|| \leq \epsilon \Longrightarrow argmax\ f(X) = c$  \\ 
		Standard robustness (SR) & 	$\forall X: ||X-\hat{X}|| \leq \epsilon \Longrightarrow ||f(X) - f(\hat{X})|| \leq \delta$ \\
		Lipschitz robustness (LR) & 	$\forall X: ||X-\hat{X}|| \leq \epsilon \Longrightarrow ||f(X) - f(\hat{X})|| \leq L||X-\hat{X}||$  \\
		Approximate CR (ACR) & 	$\forall X: ||X-\hat{X}|| \leq \epsilon \Longrightarrow f(X)_c \geq \eta$  \\
		\bottomrule
	\end{tabular}	\caption{\footnotesize{\emph{Definitions of neural network robustness~\cite{CKDKKAE22}, given a neural network $f$. The definitions assume given constant values for $\epsilon$, $\delta$, $\eta$, $L$ and some defined distance metric $|| . ||$, such as e.g.\ Euclidean distance (or $L_2$ norm) or $L_0$ norm. (Approximate) classification robustness refers to  a classifier $C$ (applied to $f$)  that will  classify $X$ as $c$ where $c$ is $\hat{X}$'s class in the input data.}}}	\label{table:robustness}
\end{figure*}

We will now apply \checkinn to reachability verification problems.
The most popular
reachability property in neural network verification is robustness.
Informally, a CNN's
robustness is its ability to correctly classify an input to which a small
perturbation is applied. More specifically, \emph{a CNN is $\epsilon$-ball
	robust for an image} if, whenever the distance between the perturbed image and
the original is no more than $\epsilon$, the CNN classifies the perturbed image
correctly.

Different techniques exist to ensure network robustness during training: data
augmentation \cite{shorten_survey_2019}, adversarial training
\cite{madry_towards_2018}, or training with logical constraints
\cite{fischer_dl2_2019}; and \cite{CKDKKAE22} shows that
these different methods give rise to different formal definitions of robustness,
which we summarise in Fig.~\ref{table:robustness}. All properties can be
written in first-order logic, and in general are amenable to SMT solvers.
Imandra can also express these properties, with the benefit of a somewhat more
intuitive syntax than the solvers admit. For example, this is \checkinn
definition of standard robustness (using $L_0$-norm distance function on
vectors):


\begin{lstlisting}[label={lst:sr}]
let sr model input delta epsilon ?(constraint=true) x = 
let y = model input in
let fx = model x in
let dist = bind2 y fx L0 in
constraint && (L0 x input) <=? epsilon ==> dist <=? delta
\end{lstlisting}

\noindent We refer the reader to \checkinn code for the remaining
three robustness definitions, which use similar syntax. We note the addition of
a parameter \lstinline{constraint} on admissible CNN inputs, which we
often use as a validity check for the type of input images that the network
accepts, as will be illustrated later in this section.

Robustness is best amenable to proofs by arithmetic manipulation. This explains the interest of the SMT-solving community in the topic, which started with using Z3 directly~\cite{HuangKWW17}, and has resulted in highly efficient SMT solvers specialised in robustness proofs for neural networks~\cite{KaBaDiJuKo17Reluplex,KatzHIJLLSTWZDK19}.

In Imandra, \lstinline{[@@blast]}, a tactic for SAT-based symbolic execution modulo
higher-order recursive functions, can be applied to these problems.
However, \lstinline{blast} currently does not support real arithmetic. This
requires us to \emph{quantise} the neural networks we use (i.e.\ convert them to
integer weights) and results in a \emph{quantised CNN library} in \checkinn.
Quantisation is a common technique in  machine learning and NN verification:
quantised neural networks take less computational resources to run, are more amenable to verification, and
often can be trained to be as accurate as floating point networks~\cite{DuncanKSL20,K18,kozlov_neural_2021}.  
Modulo this hurdle, verification of the CNNs
goes in a straightforward way, and requires just one line of code. For example,
for standard robustness, this line looks like this:

\begin{lstlisting}
verify (fun x -> sr model input 1 epsilon
    ~constraint:(is_valid x) x) [@@blast]
\end{lstlisting}


\noindent Note that the code includes the validity check for input images; for
example, we may require that all input matrices are of size $9 \times 9$ and have
binary inputs. This reduces the search space and gives more tractable results.
This is also the first instance when we use the tactic syntax \lstinline{[@@blast]}.
Imandra's mode of interaction is by supplying proof details and hints to
the user, and taking additional lemmas and tactics like \lstinline{[@@blast]} as
input. In this case, just calling \lstinline{[@@blast]} completes the proof.

To illustrate the usual pattern of robustness verification, we select images
from the data set, for example those shown in Fig.~\ref{fig:dataset_sample}
and use the module that holds all verification calls as in the code above. We obtain
the results shown in Fig.~\ref{table:robustness_verification}. We can see that
all the properties terminated and Imandra gives a ``proved" or ``refuted"
result. In the latter case, Imandra gives an executable counterexample which is a benefit of the language.

The execution times given in  Fig.~\ref{table:robustness_verification} are reasonable,
but the example network and images are rather small. Already (a quantised version of) the ACAS Xu challenge~\cite{KaBaDiJuKo17Reluplex} 
is out of reach for \lstinline{[@@blast]}, which we will try to repair in the next section.   

Our conclusions are two-fold. Firstly, we notice the payoff
of implementing a large general library for CNNs: we can now implement and
verify robustness properties in just a few lines of code, in a clear syntax.
This shows Imandra's ease of use as a verification tool.
Secondly, we managed to experimentally confirm the suggestion by
\cite{CKDKKAE22} that verifying different definitions of
robustness on the same network yields different results; this speaks for the
importance of distinguishing between formal definitions of robustness, and for
the future usability of our Imandra library that provides all these definitions
in a generic way. And finally, this points to a future research direction --
connecting Imandra with neural network-specific solvers like Marabou, in which case a call
of a procedure similar to \lstinline{[@@blast]} could perhaps deal with the queries
more efficiently; moreover, it would open the way for such proofs in the
real-valued version of our CNN library.


\begin{figure}
	\newcommand\nocell[1]{\multicolumn{#1}{c|}{}}
	\begin{tabular}{ c  c  c  c  c }
		\toprule 
		\multicolumn{1}{c}{} &
		\multicolumn{2}{c}{\textbf{Happy}} & 
		\multicolumn{2}{c}{\textbf{Sad}} \\
		\cmidrule(rl){2-3}\cmidrule(rl){4-5}
		\textbf{Property} & Result & Time (s) & Result & Time (s) \\
		\midrule
		$CR(\epsilon)$ & Refuted & 96.36 & Refuted & 89.35 \\
		$SR(\epsilon, \delta)$ & Proved & 107.12 & Proved & 108.37 \\ 
		$LR(\epsilon, L)$ & Proved & 110.74 & Proved & 117.43 \\ 
		$ACR(\epsilon, \eta)$ & Refuted & 90.47 & Proved & 89.00\\ 
		\bottomrule
	\end{tabular}
		\caption[CNN robustness verification results.]{\emph{\footnotesize{CNN robustness verification results, for $\epsilon$-balls in the vicinity of the two given images. The parameters are: $\epsilon=1,\delta=1,L=2,\eta=1$.}}}	\label{table:robustness_verification}
\end{figure}
\section{Matrix Representations}\label{sec:acas}


In this section, we address limitations discovered in the previous section, and extend \checkinn to construct reachability proofs with real numbers and larger networks. In particular, 
we put to the test Imandra's native automation procedures. 
We take the ACAS Xu
verification benchmark~\cite{KaBaDiJuKo17Reluplex}. An ACAS Xu neural network has inputs that model an input state of an aeroplane composed of five components:
distance  from  ownship  to an intruder, angle  to the intruder, speed of ownship (\lstinline{vown}), speed of the intruder (\lstinline{vint}).  

%
The network's output is a vector whose elements represent actions:  clear-of-conflict (COC), weak right, strong right, weak left, or strong left. In line with Section~\ref{sec:background},
a function like $argmax$ can classify the network's input into one of these classes based on the top value in the network's output. In~\cite{KaBaDiJuKo17Reluplex}, we find 45 ACAS Xu networks;   each has 5-6 layers, of up to a dozen of neurons in each, all layers are fully connected and have $relu$ activation functions ($relu(x) = x \textrm{ if } x\geq 0 \textrm{ else } 0$). 

%
Ten safety properties are defined for these networks in \cite{KaBaDiJuKo17Reluplex}.
For example, property $\phi_1$ states: ``If the intruder is distant and is significantly slower than the ownership, the score of a COC advisory will always be below a certain fixed threshold". 
Taking specific constants from~\cite{KaBaDiJuKo17Reluplex}, the left side of  the implication can be defined in Imandra as:
\begin{lstlisting}
	let condition1 (dist, vown, vint) =
	(dist >= 55948) && (vown >= 1145) && (vint <= 60)
\end{lstlisting}


\noindent Similarly to robustness properties, this verification property could be handled by general-purpose SMT solvers; however, as~\cite{KaBaDiJuKo17Reluplex} points out, they do not scale.
Indeed, when we use the \checkinn on quantised ACAS Xu neural networks, \lstinline{[@blast]} does not terminate. 
This is why the algorithm \emph{Reluplex} was introduced in \cite{KaBaDiJuKo17Reluplex} as an additional heuristic to SMT solver algorithms;
\emph{Reluplex} has since given rise to a domain specific solver \emph{Marabou}~\cite{KatzHIJLLSTWZDK19}.

\subsection{Leveraging Imandra's Native Automation: Matrices as Functions}\label{sec:maps}

We start with keeping the integer values for weights but redefining matrices as functions (from indices to values), which gives
constant-time (recursion-free) access to matrix elements:

\begin{lstlisting}
	type arg =
	| Rows
	| Cols
	| Value of int * int
	
	type 'a t = arg -> 'a
	
let nth (m: 'a t) (i: int) (j: int): 'a = m (Value (i,j))
\end{lstlisting}

Note the use of the \lstinline{arg} type, which treats a matrix as a function evaluating ``queries'' (e.g., ``how many rows does this matrix have?'' or ``what is the value at index $(i,j)$?''). This formalisation technique is used as Imandra's logic does not allow function values inside of algebraic data types. 
We thus recover some functionality given by refinement types in~\cite{KokkeKKAA20}.  

Furthermore, we can map over a matrix, map2 over a pair of matrices, transpose a
matrix, construct a diagonal matrix etc.\ without any recursion, since we work
point-wise on the elements. At the same time, we remove the need for error
tracking to ensure matrices are of the correct size: because our matrices are
total functions, they are defined everywhere (even outside of their stated
dimensions), and we can make the convention that all matrices we build are valid
and sparse by construction (with default 0 outside of their dimension bounds).

For full definitions of matrix operations and layers, the reader is referred to \checkinn, but we will give some definitions here, mainly to convey the general style (and simplicity!) of the code.  A script transforms the original ACAS Xu networks into a sparse functional matrix representation.
For example, layer 5 of one of the networks we used is defined as follows (\lstinline{fc} stands for a fully-connected layer):


\begin{lstlisting}
	let layer5 = fc relu (
	function
	| Rows -> 50
	| Cols -> 51
	| Value (i,j) -> Map.get (i,j) layer5_map)
	
	let layer5_map =
	Map.add (0,0) (1) @@
	Map.add (0,10) (-1) @@
	Map.add (0,29) (-1) @@
	...
	Map.const 0
\end{lstlisting}

\noindent Networks are compressed in order to reduce the number of computations using two well-known compression methods. On one hand, they are quantised, i.e. the real-valued weights are converted into integers using \textit{static quantisation}~\cite{K18}. On the other hand, the weights are pruned using magnitude as a pruning criterion, meaning that weights with the lowest absolute value are removed.

\noindent
We can model the resulting neural network via a function \lstinline{run}:

\begin{lstlisting}
let run (dist, angle, angle_int, vown, vint) =
let m = mk_input (dist, angle, angle_int, vown, vint) in
layer0 m |> layer1 |> layer2 |> layer3 |> layer4 |> layer5 |> layer6
\end{lstlisting}

\noindent  Note that we no longer need to use monadic binds, as we no longer track dimension errors. We can now define the first ACAS Xu property~\cite{KaBaDiJuKo17Reluplex}: 

\begin{lstlisting}
	let property1 x =
	let output = run x in
	let coc = Matrix.nth output 0 0 in
	coc <= 1500
	
	theorem acas_xu_phi_1 x =
	is_valid x && condition1 x ==> property1 x
\end{lstlisting}

\noindent The only help Imandra needs to prove this automatically are the forward-chaining rules about the $relu$ function:

\begin{lstlisting}
	lemma relu_pos x =
	x >= 0 ==> (relu x) [@trigger] = x
	[@@auto] [@@fc]
	
	lemma relu_neg x =
	x <= 0 ==> (relu x) [@trigger] = 0
	[@@auto] [@@fc]
\end{lstlisting}

\noindent
And then we disable $relu$ expansion for all of the proofs using the \lstinline{[@@disable]} annotation. This way,
$relu$ induces no simplification case-splits, while all
relevant information about $relu$ values is propagated,
per instance, on demand to our simplification context.
Now Imandra's engine takes care of the proof automatically  (when we use the tactic \lstinline{[@@auto]}), and takes just under 1.5 minutes. 
In Figure~\ref{tab:acas} we give a representative evaluation of Imandra's performance on several ACAS Xu networks and properties. We set the timeout time to 5 minutes, and approximately half of the cases terminate within the time limit. Execution time is orders of magnitude faster than Marabou's time on full ACAS Xu networks, which may suggest that combining pruning and verification~\cite{LahavK21} is a good direction for Imandra. We leave a thorough investigation of this for future work. 

Several factors played a role in automating the proofs. Firstly, Imandra being a
higher-order functional language opened the way for us to experiment with
alternative matrix representations in the first place. By using maps for the
large matrices, we eliminate all recursion (and large case-splits) except for
matrix folds (which now come in only via the dot product), which allowed Imandra
to expand the recursive matrix computations ``on demand.'' 
Finally, Imandra's native simplifier contributed to the success. It works on a DAG representation of terms and speculatively expands instances of recursive functions, only as they are (heuristically seen to be) needed.
Incremental congruence closure and simplex data structures are shared across DAG nodes, and symbolic execution results are memoised. Moreover, forward-chaining rules (such as those characterising $relu$) are only applied on demand.
Informally speaking, Imandra works lazily expanding out the linear algebra as it is needed, and eagerly with sharing information over the DAG.
Contrast this approach with that of \lstinline{reluplex} which, informally, starts with the linear algebra fully expanded, and then works to derive laziness and sharing. 


\subsection{Extension to Reals}

Section~\ref{sec:background} defined matrices as lists of lists; and that definition
in principle worked for both integer and real-valued matrices. However, we could
not use \lstinline{[@@blast]} to automate proofs when real values were involved;
this meant we were restricted to verifying integer-valued networks. The matrix-as-function implementation can be extended to proofs with real-valued matrices; however, it is not a trivial extension. In Section~\ref{sec:maps},  the matrix's value was of
the same type as its dimensions. Thus, if the matrix elements are real-valued, then
in this representation the matrix dimensions will be real-valued as well.
But this complicates termination guarantees for functions which do recursion along
matrix dimensions.


To simplify the code and the proofs, three potential solutions were considered:\\
1. Use an algebraic data type for results of matrix queries: this introduces pattern matching in the implementation of matrix operations, which reduces proof search efficiency.\\
2. Define a matrix type with real-valued dimensions and values: this poses the problem of proving the function termination when using matrix dimensions in recursion termination conditions.\\
3. Use \emph{records} to provide polymorphism and allow matrices to use integer dimensions and real values.

In an accompanying note~\cite{DPK22} and in Appendix~\ref{ap:records}, we provide further details on each of the three implementations in \checkinn.
But the second option was a clear winner when it came to evaluating it on ACAS Xu. We therefore only highlight its features here.
The implementation is symmetric to the one using integers:



\begin{lstlisting}
type arg =
	| Rows
	| Cols
	| Value of real * real

type 'a t = arg -> 'a
\end{lstlisting}

\noindent A problem arises in recursive functions where matrix dimensions are used as decrementors in stopping conditions, for instance in the \lstinline{fold_rec} function used in the implementation of the folding operation.
Imandra only accepts definitions of functions for which it can prove termination. The dimensions being real numbers prevents Imandra from being able to prove termination without providing a custom measure. In order to define this measure, we need to connect the continuous world of reals with the discrete world of integers (and ultimately ordinals) for which we have induction principles. We chose to develop a \lstinline{floor} function that allows Imandra to prove termination with reals.


To prove termination of our \lstinline{fold_rec} function recursing along reals, we define an \lstinline{int_of_real : real -> int} function in Imandra, using a subsidiary \lstinline{floor : real -> int -> int} which computes an integer floor of a real by ``counting up'' using its integer argument. In fact, as matrices have non-negative dimensions, it suffices to only consider this conversion for non-negative reals, and we formalise only this. We then have to prove some subsidiary lemmas about the arithmetic of real-to-integer conversion, such as:
\begin{lstlisting}
lemma floor_mono x y b =
  Real.(x <= y && x >= 0. && y >= 0.)
  ==> floor x b <= floor y b

lemma inc_by_one_bigger_conv x =
Real.(x >= 0. ==> int_of_real (x + 1.0) > int_of_real x)
\end{lstlisting}

\noindent Armed with these results, we can then prove termination of \lstinline{fold_rec} and admit it into Imandra's logic via the ordinal pair measure below:
\begin{lstlisting}
[@@measure Ordinal.pair
            (Ordinal.of_int (int_of_real i))
            (Ordinal.of_int (int_of_real j))]
\end{lstlisting}

\begin{figure*}
	\caption{Results of experiments ran on the properties and networks from the ACAS Xu benchmark~\cite{katz_reluplex_2017}. The verifications were run on virtual machines with four 2.6 GHz Intel Ice Lake virtual processors and 16GB RAM. Timeout was set at 5 hours}\label{tab:acas}
	\centering
	\begin{tabular}{{ P{1.2cm} P{1.7cm} P{1.7cm} P{1.3cm} P{1.7cm} P{1.3cm} P{1.7cm} P{1.3cm} }}
		\toprule
		
		\multicolumn{2}{P{3.3cm}}{} &
		\multicolumn{2}{P{3.1cm}}{\textbf{CheckINN: Pruned and Quantised Networks}} &
		\multicolumn{2}{P{3.1cm}}{\textbf{CheckINN: Pruned Networks}} &
		\multicolumn{2}{P{3.6cm}}{\textbf{Reluplex: Full ACAS Xu Networks}} \\
		
		\cmidrule(rl){3-4} \cmidrule(rl){5-6} \cmidrule(rl){7-8}
		
		\textbf{Property} & \textbf{Result} & Quantity & Time (s) & Quantity & Time (s) & Quantity & Time (s) \\ 
		
		\rowcolor{lavender}
		$\phi1$ & SAT 		& 4  & 258	& 20 & 13387 	& 0  &			\\
		\rowcolor{lavender}
		& UNSAT 	& 0  & 		& 0	 & 		 	& 41 & 394517	\\
		\rowcolor{lavender}
		& TIMEOUT 	& 5  & 		& 24 & 			& 4  &			\\
		
		$\phi2$ & SAT 		&    & 		& 7  & 2098 	& 35 & 82419	\\
		& UNSAT 	& 0  & 		& 2	 & 896	 	& 1  & 463		\\
		& TIMEOUT 	&    & 		& 26 & 			& 4  &			\\
		
		\rowcolor{lavender}
		$\phi3$ & SAT 		&    & 		& 39 & 10453 	& 35 & 82419	\\
		\rowcolor{lavender}
		& UNSAT 	& 0  & 		& 0	 & 		 	& 1  & 463		\\
		\rowcolor{lavender}
		& TIMEOUT 	&    & 		& 2  & 			& 4  &			\\
		
		$\phi4$ & SAT 		& 16 & 1422	& 36 & 21533	& 0  & 		 	\\
		& UNSAT 	& 1  & 114	& 0  & 			& 32 & 12475 	\\
		& TIMEOUT 	& 18 & 3	& 5  & 			& 0  &			\\
		
		\rowcolor{lavender}
		$\phi5$ & SAT 		& 1  & 57	& 1  & 98	 	& 0  & 			\\
		\rowcolor{lavender}
		& UNSAT 	& 0  &		& 0  & 		 	& 1	 & 19355	\\
		
		$\phi6$ & SAT 		& 1  & 196  & 1  & 98		& 0  &			\\
		& UNSAT 	& 0  & 		& 0  & 			& 1  & 180288	\\
		
		\rowcolor{lavender}
		$\phi7$ & TIMEOUT 	& 1	 &		& 1  &	 		& 1  & 			\\
		
		$\phi8$ & SAT 		& 0  & 		& 0  & 			& 1  & 40102 	\\
		& TIMEOUT 	& 1  & 		& 1  & 			& 0  & 			\\
		
		\rowcolor{lavender}
		$\phi9$ & SAT		& 1  & 66	& 1  & 109	 	& 0  & 			\\
		\rowcolor{lavender}
		& UNSAT		& 0  &		& 0  & 			& 1  & 99634 	\\
		
		$\phi10$ & SAT		& 1 & 116	& 0  &			& 0  &			\\
		& UNSAT 	& 0 & 		& 1  & 637 		& 1  & 19944	\\
		& TIMEOUT 	& 0 & 		& 0  &	 		& 0 &			\\
		\bottomrule
	\end{tabular}
\end{figure*}

Extending the functional matrix implementation to reals was not trivial, but it did have a real payoff.
Using this representation, we were able to verify real-valued versions of the pruned ACAS Xu networks!
In both cases of integer and real-valued matrices, we pruned the networks to 10\% of their original size.
So, we still do not scale to the full ACAS Xu challenge.  However, the positive news is that the real-valued version
of the proofs
uses the same waterfall proof tactic of Imandra, and requires no extra effort from the programmer to complete the proof.
Moreover, as preliminary evaluation in Figure~\ref{tab:acas} shows, the real values do not substantially increase verification times.
This result is significant bearing in mind that many functional and higher-order theorem provers are known to have significant drawbacks when switching to real numbers.

\section{Conclusions, Related and Future Work}\label{sec:concl}

\subsection{Paper Summary} 
 \checkinn defined, as broadly as possible, the design space for neural
network verification in ITP. As far as we know, no other single
existing tool~\cite{KatzHIJLLSTWZDK19,HuangKWW17,SinghGPV19,GeMiDrTsChVe18,AEHW20} or library~\cite{TassarottiV0T21,MariaBLFGRG22,BS19} has yet managed to cover such a wide range of verification tasks. 
We have taken advantage of both the wide range of choices for matrix representation available in Imandra when it came to reachability proofs,
and the facility to combine first-order and higher-order object definitions, proofs by induction, simplification and decision procedures in the structural proofs.

The resulting \checkinn library contains several parts: 
\begin{enumerate}
\item[a)] a set of modules that use a monadic implementation of matrices as lists that includes a CNN and an FNN implementation,
  both of which support quantised and real-valued networks (this library was used in Sections~\ref{sec:background}, \ref{sec:CNN} and \ref{sec:verify});
  	\item[b)] a library with several alternative abstract definitions of FNNs specifically tailored for reasoning by induction (used in Section~\ref{sec:induct});
	\item[c)]  a library for FNNs using matrices implemented as functions and records for both real- and integer-valued networks (used in Section~\ref{sec:acas}).
\end{enumerate}

They are not currently connected, but complement each other as follows. 
Part a) is best suited to implement and run CNNs that are compiled directly from a Python environment; and it works reasonably well for robustness verification for small quantised networks.
Part b) 
is built to yield automated inductive proofs that exploit structural properties of FNNs.
Part c) performs particularly well with reachability analysis for larger FNNs and supports real numbers. 

\subsection{Contributions with Respect to Related Work}
CNN formalisation and formulation of structural properties of CNN are both original contributions. We are not aware of any prior similar results in any ITP. 

\textbf{Matrix Representations.} We showed that the choice of matrix representation favours certain kinds of proofs. Matrices as lists are well-amenable to structural proofs by induction, while matrices as functions or records help to scale reachability proofs. Flexibility with matrix choices proved to be a useful feature.
Real numbers in Imandra allowed for smooth transitions from integer to real parts of the library, especially in inductive proofs.

\noindent FP literature gives a selection of different matrix representation methods. Matrices as lists  are considered in \cite{heras_incidence_2011} (in the context of dependent types in Coq), in ~\cite{KokkeKKAA20} (in the context of refinement types of F$^*$) and in \cite{grant_sparse_1996} (for sparse matrix encodings in Haskell).
The difference between the list and function approaches was
discussed in~\cite{wood_vectors_2019} (in Agda, but with no neural network application in mind).
Our main contribution here is to trace the connection between matrix representations and the automation of different kinds of proofs. 
We believe that the methods we described could be useful in other theorem provers (both first- and higher-order) that combine functional programming and automated proof methods, such as ACL2, PVS, Isabelle/HOL and Coq. 
For example, in all these systems functions defining matrix operations (e.g., convolution) over lists  are often more complex compared to their counterparts over matrices represented as functions, which can benefit from non-recursive definitions.
Overall, as these various prominent theorem proving systems work ultimately with functional programs over algebraic datatypes, our core observations carry over to them in a natural way.

\textbf{Structural Verification of NN.} 
De Maria et al.~\cite{MariaBLFGRG22} formalise in Coq \emph{``neuronal
	archetypes''} for biological neurons. Each archetype is a specialised kind of
perceptron (a small, typically single-unit, neural network), in which additional functions are added to amplify or inhibit the
perceptron's outputs.
The paper collects a rich variety of structural properties and proofs characterising these archetypes, formalised in Coq.
In this paper, we worked with neural networks used in classification, and unlike~\cite{MariaBLFGRG22} had to work with matrices.
Defining structural properties for such networks was more challenging. While the monotonicity proofs of
Section~\ref{sec:induct} do not differ much in their complexity from ~\cite{MariaBLFGRG22} (we may only note the greater power of inductive proof automation that Imandra offers), the proofs of Extreme Values Lemma would have been hard to replicate in Coq as simply as we did it with Imandra. In particular, the bounded model checking tactic used in that section is unique to Imandra.

\textbf{Reachability Verification of NN.}
Although Imandra's simplifier-based automation
did not scale
to the original dense ACAS Xu network verified by Reluplex~\cite{katz_reluplex_2017},
we are encouraged that the obtained proofs were achieved without tuning Imandra's generic proof automation
strategies. No other ITP we know of would be able to achieve that much using its native tactics. 
We are hopeful that the development of neural-network specific tactics 
will help Imandra
scale to bigger networks in the future. Indeed, directly connecting Imandra with Marabou or other similar solvers is also a possible future direction.

\subsection{Future Work}
Some other considerations were left for future work.
This paper draws a wide range of NN verification methods, without aiming at any single verification challenge in particular. Our next step is
to apply these methods to some significant verification task.
In this respect, extending structural verification of CNN to real-life data sets and scenarios, and further automation of reachability proofs have high priority.  

Current state of the art reachability solvers like Marabou~\cite{KatzHIJLLSTWZDK19} rely on SMT solvers, which cannot handle exponents and logarithms,
and this explains why they only work with networks with linear activation functions. Although CheckINN does not
extend state of the art in this aspect yet, we hope to apply Imandra's
existing experience with using approximating recursive functions (e.g., Cauchy sequences) to non-linear activation functions commonly used in NNs.

There is an important question of the numerical types used in
neural networks, that still awaits a successful resolution by the theorem
proving communities. We used real and integer-valued networks. Mainstream
work in Python works with floats, most SMT-based solvers use rationals or
restricted reals~\cite{KaBaDiJuKo17Reluplex,KatzHIJLLSTWZDK19}, abstract
interpretation tools can use floating points~\cite{SinghGPV19} and some ITPs are
amenable to formalisations with reals~\cite{KokkeKKAA20} and even floating
points~\cite{BS19}. But it is known that transition from one to another may
render sound proofs unsound~\cite{JiaR21}, and so the choices cannot be taken
lightly. 

This paper only addresses the problem of analysing and verifying already trained
neural networks. There may be demand for  verification of machine learning
algorithms (as was done e.g.\ in~\cite{TassarottiV0T21} for decision stumps), which is
worth exploring in the future.

Finally, recent developments in neural network verification suggest that training should
become part of verification techniques~\cite{fischer_dl2_2019,CKDKKAE22,slusarz_differentiable_2022}. Exploring Imandra's facilities
to sustain the ``training-for-verification'' cycle is left for future work.


\bibliographystyle{ACM-Reference-Format}
\bibliography{main}

\appendix
\section{Monadic Operations on Results}\label{ap:monad}

In the matrix as list implementation, we check matrix size for matrix operations. In some operations, 
such as dot product, the dimension checks are important. When matrix sizes do not match,
we use OCaml's \lstinline{result} type in order to model the error.

The result type is simply defined as

\begin{lstlisting}
	let type ('a, 'b) result = 
		| Ok of 'a
		| Error of 'b
\end{lstlisting}

The result type is accompanied by the \lstinline{bind} and \lstinline{return} functions, which make it a monad, as well as common monadic operations. As Imandra is a pure language where side-effects are not possible, the monadic operations lets us propagate errors. 

The \lstinline{bind} function allows to chain functions that may fail, i.e. functions that 
take as input a ``plain'' value and return a \lstinline{result}. If the input is an error, the error is propagated, otherwise, the function is applied to the value contained within the \lstinline{result}. It is defined as:

\begin{lstlisting}
	let bind (f: 'a -> ('b, 'c) result) (x: ('a, 'c) result) : ('b, 'c) result = function
		| Ok x'   -> f x'
		| Error e -> Error e 
\end{lstlisting}

For instance, in the matrix-as-function implementation (which does \emph{not} use \lstinline{result}), a network is formed by chaining the output of a layer as input to the next layer. For this, the standard OCaml operator \lstinline{|>} is used:

\begin{lstlisting}
	let run (dist, angle, angle_int, vown, vint) =
		let open Weights in
		let m = mk_input (dist, angle, angle_int, vown, vint) in
		layer0 m |> layer1 |> layer2 |> layer3 |> layer4 |> layer5 |> layer6
	;;
\end{lstlisting}

When using \lstinline|result|, we have to use the \lstinline{(>>=)} (bind) operator. With this operator, if an error occurs in layer1, all subsequent layers will pass this error to the next until the output:

\begin{lstlisting}
	let model input = layer_0 input >>= layer_1 >>= layer_2 >>= layer_4 >>= arg_max;;
\end{lstlisting}

From the \lstinline{bind} function and the \lstinline{return} function, we can define a series of helper functions that improve code readability and maintainability. For instance, \lstinline{lift} applies a function over  ``plain'' values to a \lstinline{result}; \lstinline{bind2} and \lstinline{lift2} are equivalents of \lstinline{bind} and \lstinline{lift} for functions that take two arguments instead of one; \lstinline{flatten} reduces nested \lstinline{result} type to a simple result type (e.g. \lstinline{Ok (Ok (Ok 2))} is simplified to \lstinline{Ok 2}), etc.

Comparison operators between \lstinline{result} are also implemented to be used in verification properties. Thus, to compare two \lstinline{(int, string) result}, the operators \lstinline{(<=?), (=?) (>=?)} are used instead of the ``plain'' operators \lstinline{(<=), (=), (>=)}

\section{Full Interaction Cycle for Inductive Proof of Monotonicity  in Imandra}\label{app:code}

The full proof of the Monotonicity Lemma, generated by Imandra, is given in Listing~\ref{lst:mon}. Of notable interest are the inductive schemes it generated and discarded, and automated proof search by simplification.  

\begin{lstlisting}[float=*, frame=single, language=caml, caption={Imandra's proof of Monotonicity Lemma, Part 1}, label={lst:mon}]
	val network_monotonicity :
	real vector vector vector ->
	real vector vector -> real vector -> real vector -> bool = <fun>
	Goal:
	
	positive_3d ws && positive_2d bs && positive i && gte i' i
	==> gte (network ws bs i') (network ws bs i).
	
	1 nontautological subgoal.
	
	Subgoal 1:
	
	H0. positive_3d ws
	H1. positive_2d bs
	H2. positive i
	H3. gte i' i
	|---------------------------------------------------
	gte (network ws bs i') (network ws bs i)
	
	Must try induction.
	
	The recursive terms in the conjecture suggest 6 inductions.
	Subsumption and merging reduces this to 2.
	
	However, scheme scoring gives us a clear winner.
	We shall induct according to a scheme derived from network.
	
	Induction scheme:
	
	(not (ws <> [] && bs <> []) ==> phi bs i i' ws)
	&& (bs <> []
	&& ws <> []
	&& phi (List.tl bs) (layer (List.hd ws) (List.hd bs) i)
	(layer (List.hd ws) (List.hd bs) i') (List.tl ws)
	==> phi bs i i' ws).
	
	2 nontautological subgoals.
	
	Subgoal 1.2:
	
	H0. positive_3d ws
	H1. positive_2d bs
	H2. positive i
	H3. gte i' i
	|---------------------------------------------------
	C0. ws <> [] && bs <> []
	C1. gte (network ws bs i') (network ws bs i)
	
	But simplification reduces this to true, using the definition of network.
	
	Subgoal 1.1:
	
	H0. positive_3d ws
	H1. positive_2d bs
	H2. positive i
	H3. gte i' i
	H4. bs <> []
	H5. ws <> []
	H6. ((positive_3d (List.tl ws) && positive_2d (List.tl bs))
	&& positive (layer (List.hd ws) (List.hd bs) i))
	&& gte (layer (List.hd ws) (List.hd bs) i')
	(layer (List.hd ws) (List.hd bs) i)
	==> gte
	(network (List.tl ws) (List.tl bs)
	(layer (List.hd ws) (List.hd bs) i'))
	(network (List.tl ws) (List.tl bs)
	(layer (List.hd ws) (List.hd bs) i))
	|---------------------------------------------------
	gte (network ws bs i') (network ws bs i)
	
	But simplification reduces this to true, using the definitions of network,
	positive_2d and positive_3d, and the rewrite rules layer_monotonicity and
	positive_push_2d.
	
	Rules:
	(:def network)
	(:def positive_2d)
	(:def positive_3d)
	(:rw layer_monotonicity)
	(:rw positive_push_2d)
	(:fc gte_preservation)
	(:fc pos_tl_2d)
	(:induct network)
	
	Theorem proved.
\end{lstlisting}

\section{Proof of the Extreme Value Theorem}\label{ap:extreme}

\subsection{Manual Proofs}

In this section, we provide proofs for the Extreme Values Lemma presented in the main text~\ref{lem:extreme} in the two specific cases \emph{R1} and \emph{R2}. We re-use the same notation as in the lemma's definition(cf. Section~\ref{subsec:structural_property}).

\begin{lemma}[Extreme Values]
	Let $a$ be a vector $a_1, \ldots , a_n$ with a distinct pattern and extreme values $a_1, \ldots , a_m$. If for  $w^x$ and $w^y$, we have 
	\begin{equation}\label{eq:extr_val0}
	w^x_1 > w^y_1 \land \ldots \land w^x_m > w^y_m,
	\end{equation}
	
	then 
	\begin{equation}\label{eq:extr_val1}
	(a_1  w^x_1 + \ldots + a_n  w^x_n) > (a_1  w^y_1 + \ldots + a_n w^y_n) .
	\end{equation}
	
\end{lemma} 

\subsubsection{Case 1: Binary Pooling Layer Output}
Condition \textbf{R1}: $a$ is a binary vector, and $a_{min} \neq a^*, a_{max} \neq a^*$ 

For $a_1, ..., a_m \in a_{ex}$ we need to have 
$$a_1 =\ldots = a_m = 1$$
as
$a^* \neq a_{max} \neq a_{min}$

For $a_{m+1}, \ldots, a_n$ we need to have 
$$a_{m+1} = \ldots = a_n = 0$$ 
as $a^* \neq a_{max} \neq a_{min}$.

Then we have 
$$w_1^x + \ldots + w_m^y > w_1^y + \ldots + w_m^y$$
as (\ref{eq:extr_val1}). This holds because by (\ref{eq:extr_val0}), 
$w_i^x > w_i^y$ for $i \in [1,m]$.

\subsubsection{Case 2: Binary Pooling Layer Output}

Condition \textbf{R2}: $w_x$ and $w_y$ are binary vectors, $a^* \neq 0$

We suppose that $a_1, \ldots, a_m$ are the smallest possible, i.e. close to $a^* + \frac{a_{max} - a^*}{2}$
and $a_{m+1}, \ldots, a_n$ are the greatest possible, i.e. close to $a^*$

since $w^x$ and $w^y$ are binary, and $w_1^x > w_1^y, \ldots, w_m^x > w_m^y$, we know that 
$$ w_1^x = \ldots = w_m^x = 1$$
$$ w_1^y = \ldots = w_m^y = 0$$

We need to prove that
\begin{align*}
m \left(a^* + \frac{a_{max} - a^*}{2}\right) &> (n-m)a^* \\
1.5 m a^* + \frac{m a_{max}}{2} &> na^* \\
1.5 m + \frac{m a_{max}}{2a^*} &> n \\
m\left(1.5 +\frac{a_{max}}{2a^*}\right) &> n
\end{align*}

We know that $m < n$, so the condition is $m$ has to be \emph{at least} $\frac{n}{1.5 + \frac{a_{max}}{2a^*}}$

\subsection{Details of Imandra proofs}

\subsubsection{Extreme Values Lemma -- Case R1}
We provide full proof produced automatically by Imandra. We split it into Listings~\ref{lst:ev1}, \ref{lst:ev2}, \ref{lst:ev3} and \ref{lst:ev4}.
Of notable interest is the automatically generated inductive scheme, as well as auxiliary lemmas that completed the proof. 

\begin{lstlisting}[float=*, frame=single, language=caml, caption={Imandra's proof of Extreme Value Lemma, Part 1}, label={lst:ev1}]
# lemma main vs =
    is_valid_r1 vs
    ==>
    let (p1,p2) = dot_products vs in
    p1 >. p2
  [@@auto]
  ;;
val main : value list -> bool = <fun>
Goal:

is_valid_r1 vs ==> let (p1, p2) = dot_products vs in p1 >. p2.

1 nontautological subgoal.

Subgoal 1:

 H0. rec_fun.is_valid_r1.aux.0 false vs
 H1. (dot_products vs).0 <=. (dot_products vs).1
|---------------------------------------------------------------------------
 false


Must try induction.

The recursive terms in the conjecture suggest 2 inductions.
Subsumption and merging reduces this to 1.

We shall induct according to a scheme derived from dot_products.

Induction scheme:

 (not (not Is_a(Extreme, List.hd vs) && vs <> [])
  && not (Is_a(Extreme, List.hd vs) && vs <> []) ==> φ vs)
 && (vs <> []
     && Is_a(Extreme, List.hd vs) && φ (List.tl vs) && φ (List.tl vs)
     ==> φ vs)
    && (vs <> []
        && not Is_a(Extreme, List.hd vs)
           && φ (List.tl vs) && φ (List.tl vs)
        ==> φ vs).

3 nontautological subgoals.

Subgoal 1.3:

 H0. rec_fun.is_valid_r1.aux.0 false vs
 H1. (dot_products vs).0 <=. (dot_products vs).1
 H2. not (not Is_a(Extreme, List.hd vs) && vs <> [])
 H3. not (Is_a(Extreme, List.hd vs) && vs <> [])
|---------------------------------------------------------------------------
 false

But simplification reduces this to true, using the definitions of
dot_products and rec_fun.is_valid_r1.aux.0.

Subgoal 1.2:

 H0. rec_fun.is_valid_r1.aux.0 false vs
 H1. (dot_products vs).0 <=. (dot_products vs).1
 H2. vs <> []
 H3. Is_a(Extreme, List.hd vs)
 H4. not (rec_fun.is_valid_r1.aux.0 false (List.tl vs))
     || not ((dot_products (List.tl vs)).0 <=. (dot_products (List.tl vs)).1)
|---------------------------------------------------------------------------
 false

This simplifies, using the definitions of dot_products and
rec_fun.is_valid_r1.aux.0 to:

\end{lstlisting}

\begin{lstlisting}[float=*, frame=single, language=caml, caption={Imandra's proof of Extreme Value Lemma, Part 2}, label={lst:ev2}]


Subgoal 1.2':

 H0. ((Destruct(Extreme, 0, List.hd vs)).0
      *. (Destruct(Extreme, 0, List.hd vs)).2
      +. (dot_products (List.tl vs)).0)
     <=.
     ((Destruct(Extreme, 0, List.hd vs)).1
      *. (Destruct(Extreme, 0, List.hd vs)).2
      +. (dot_products (List.tl vs)).1)
 H1. Is_a(Extreme, List.hd vs)
 H2. rec_fun.is_valid_r1.aux.0 false vs
 H3. vs <> []
|---------------------------------------------------------------------------
 rec_fun.is_valid_r1.aux.0 false (List.tl vs)


We can eliminate destructors by the following substitution:
 vs -> vs1 :: vs2

This produces the modified subgoal:

Subgoal 1.2'':

 H0. rec_fun.is_valid_r1.aux.0 false (vs1 :: vs2)
 H1. Is_a(Extreme, vs1)
 H2. ((Destruct(Extreme, 0, vs1)).0 *. (Destruct(Extreme, 0, vs1)).2
      +. (dot_products vs2).0)
     <=.
     ((Destruct(Extreme, 0, vs1)).1 *. (Destruct(Extreme, 0, vs1)).2
      +. (dot_products vs2).1)
|---------------------------------------------------------------------------
 rec_fun.is_valid_r1.aux.0 false vs2

This simplifies, using the definition of rec_fun.is_valid_r1.aux.0 to:

Subgoal 1.2''':

 H0. rec_fun.is_valid_r1.aux.0 true vs2
 H1. (Destruct(Extreme, 0, vs1)).2 = 1
 H2. Is_a(Extreme, vs1)
 H3. ((Destruct(Extreme, 0, vs1)).0 *. (Destruct(Extreme, 0, vs1)).2
      +. (dot_products vs2).0)
     <=.
     ((Destruct(Extreme, 0, vs1)).1 *. (Destruct(Extreme, 0, vs1)).2
      +. (dot_products vs2).1)
|---------------------------------------------------------------------------
 C0. (Destruct(Extreme, 0, vs1)).0 <=. (Destruct(Extreme, 0, vs1)).1
 C1. rec_fun.is_valid_r1.aux.0 false vs2

This further simplifies to:

Subgoal 1.2'''':

 H0. rec_fun.is_valid_r1.aux.0 true vs2
 H1. (Destruct(Extreme, 0, vs1)).2 = 1
 H2. ((Destruct(Extreme, 0, vs1)).0 +. (dot_products vs2).0) <=.
     ((Destruct(Extreme, 0, vs1)).1 +. (dot_products vs2).1)
 H3. Is_a(Extreme, vs1)
|---------------------------------------------------------------------------
 C0. (Destruct(Extreme, 0, vs1)).0 <=. (Destruct(Extreme, 0, vs1)).1
 C1. rec_fun.is_valid_r1.aux.0 false vs2


We can eliminate destructors by the following
substitution:
 vs1 -> Extreme vs11

 This produces the modified subgoal:

 \end{lstlisting}
 \begin{lstlisting}[float=*, frame=single, language=caml, caption={Imandra's proof of Extreme Value Lemma, Part 3}, label={lst:ev3}]

Subgoal 1.2''''':

 H0. rec_fun.is_valid_r1.aux.0 true vs2
 H1. vs11.2 = 1
 H2. (vs11.0 +. (dot_products vs2).0) <=. (vs11.1 +. (dot_products vs2).1)
|---------------------------------------------------------------------------
 C0. vs11.0 <=. vs11.1
 C1. rec_fun.is_valid_r1.aux.0 false vs2


Must try induction.

The recursive terms in the conjecture suggest 3 inductions.
Subsumption and merging reduces this to 1.

We shall induct according to a scheme derived from dot_products.

Induction scheme:

 (not (not Is_a(Extreme, List.hd vs2) && vs2 <> [])
  && not (Is_a(Extreme, List.hd vs2) && vs2 <> []) ==> φ vs11 vs2)
 && (vs2 <> []
     && Is_a(Extreme, List.hd vs2)
        && φ vs11 (List.tl vs2) && φ vs11 (List.tl vs2)
     ==> φ vs11 vs2)
    && (vs2 <> []
        && not Is_a(Extreme, List.hd vs2)
           && φ vs11 (List.tl vs2) && φ vs11 (List.tl vs2)
        ==> φ vs11 vs2).

3 nontautological subgoals.

Subgoal 1.2'''''.3:

 H0. rec_fun.is_valid_r1.aux.0 true vs2
 H1. vs11.2 = 1
 H2. not (not Is_a(Extreme, List.hd vs2) && vs2 <> [])
 H3. not (Is_a(Extreme, List.hd vs2) && vs2 <> [])
 H4. (vs11.0 +. (dot_products vs2).0) <=. (vs11.1 +. (dot_products vs2).1)
|---------------------------------------------------------------------------
 C0. vs11.0 <=. vs11.1
 C1. rec_fun.is_valid_r1.aux.0 false vs2

But simplification reduces this to true, using the definitions of
dot_products and rec_fun.is_valid_r1.aux.0.

Subgoal 1.2'''''.2:

 H0. rec_fun.is_valid_r1.aux.0 true vs2
 H1. vs11.2 = 1
 H2. vs2 <> []
 H3. Is_a(Extreme, List.hd vs2)
 H4. (((not (vs11.2 = 1)
        || not
           ((vs11.0 +. (dot_products (List.tl vs2)).0) <=.
            (vs11.1 +. (dot_products (List.tl vs2)).1)))
       || rec_fun.is_valid_r1.aux.0 false (List.tl vs2))
      || vs11.0 <=. vs11.1)
     || not (rec_fun.is_valid_r1.aux.0 true (List.tl vs2))
 H5. (vs11.0 +. (dot_products vs2).0) <=. (vs11.1 +. (dot_products vs2).1)
|---------------------------------------------------------------------------
 C0. vs11.0 <=. vs11.1
 C1. rec_fun.is_valid_r1.aux.0 false vs2

But simplification reduces this to true, using the definitions of
dot_products and rec_fun.is_valid_r1.aux.0.

  \end{lstlisting}

\begin{lstlisting}[float=*, frame=single, language=caml, caption={Imandra's proof of Extreme Value Lemma, Part 4}, label={lst:ev4}]


Subgoal 1.2'''''.1:

 H0. rec_fun.is_valid_r1.aux.0 true vs2
 H1. vs2 <> []
 H2. not Is_a(Extreme, List.hd vs2)
 H3. (((not (vs11.2 = 1)
        || not
           ((vs11.0 +. (dot_products (List.tl vs2)).0) <=.
            (vs11.1 +. (dot_products (List.tl vs2)).1)))
       || rec_fun.is_valid_r1.aux.0 false (List.tl vs2))
      || vs11.0 <=. vs11.1)
     || not (rec_fun.is_valid_r1.aux.0 true (List.tl vs2))
 H4. vs11.2 = 1
 H5. (vs11.0 +. (dot_products vs2).0) <=. (vs11.1 +. (dot_products vs2).1)
|---------------------------------------------------------------------------
 C0. vs11.0 <=. vs11.1
 C1. rec_fun.is_valid_r1.aux.0 false vs2

But simplification reduces this to true, using the definitions of
dot_products and rec_fun.is_valid_r1.aux.0.

Subgoal 1.1:

 H0. rec_fun.is_valid_r1.aux.0 false vs
 H1. (dot_products vs).0 <=. (dot_products vs).1
 H2. vs <> []
 H3. not Is_a(Extreme, List.hd vs)
 H4. not (rec_fun.is_valid_r1.aux.0 false (List.tl vs))
     || not ((dot_products (List.tl vs)).0 <=. (dot_products (List.tl vs)).1)
|---------------------------------------------------------------------------
 false

But simplification reduces this to true, using the definitions of
dot_products and rec_fun.is_valid_r1.aux.0.

    Rules:
    (:def dot_products)
    (:def rec_fun.is_valid_r1.aux.0)
    (:induct dot_products)

   Theorem proved.
 \end{lstlisting}

 \subsubsection{Extreme Values Lemma -- Case R2}

This case needs a full formalisation of all lemma pre-conditions, as follows:


\begin{lstlisting}[ language=caml]
let rec dot_product l1 l2 = match (l1, l2) with
  | ([], _) | (_, []) -> 0.
  | (h1::t1, h2::t2) -> (h1 *. h2) +. (dot_product t1 t2)

let rec len_real = function
  | [] -> 0.
  | hd::tl -> 1. +. (len_real tl)

let rec sum = function
  | [] -> 0.
  | hd::tl -> hd +. (sum tl)

let mean l = (sum l) /. (len_real l)

let rec max' l max_val = match l with
  | [] -> max_val
  | (hd::tl) -> let max_val' = if hd >. max_val then hd else max_val in
     max' tl max_val'

let rmax (l: real list): real = match l with 
  | []      -> 0.
  | (hd::tl) -> max' tl hd

let rec min' l min_val = match l with
  | [] -> min_val
  | (hd::tl) -> let min_val' = if hd <. min_val then hd else min_val in
    min' tl min_val'

let rmin (l: real list): real = match l with
  | []      -> 0.
  | (hd::tl) -> min' tl hd

let extreme_threshold l =
  let l_max = rmax l in
  let l_mean = mean l in
  l_mean +. ((l_max -. l_mean) /. 2.)

let rec num_extreme xs e_t =
  match xs with
  | [] -> 0.
  | x::xs ->
    (if x >. e_t then 1. else 0.)
    +. num_extreme xs e_t
\end{lstlisting}

They are then assembled into R2-specific conditions:
\begin{lstlisting}[ language=caml]
  let rec r2_properties xs ys a_vals ex_threshold mean_a =
    let open Real in
    match xs, ys, a_vals with
    | [], [], [] -> true
    | x::xs, y::ys, a::a_vals ->
    (* x>y for extreme indices *)
    (a > ex_threshold ==> x>y)
    (* a is non-negative*)
    && a >=. 0.
    (* w_x and w_y are binary vectors *)
    && (x = 0. || x = 1.)
    && (y = 0. || y = 1.)
    (* a has a distinct pattern *)
    && (a > ex_threshold || a < mean_a)
    && r2_properties xs ys a_vals ex_threshold mean_a
    | _ -> false

  let extreme_value_precondition_r2 w_x w_y a =
    let mean_a = mean a in
    let e_t_a = extreme_threshold a in
    let m = num_extreme a e_t_a in
    let n = len_real a in
    let a_max = rmax a in
    r2_properties w_x w_y a e_t_a mean_a
    && Real.(m >= (n / (1.5 + (a_max / (2.0 * mean_a)))))
    && mean_a <> 0.

  let extreme_value_postcondition w_x w_y a =
    (dot_product w_x a) >. (dot_product w_y a)
\end{lstlisting}

The fully general version of the lemma (for lists of any length) could not be produced by Imandra automatically.
But taking any bounded case (a list of given finite size) allows Imandra's unrolling procedure to turn the goal into, effectively, an SMT-solving task.
This is achieved by using the tactic \lstinline{[@@unroll]}. For example:
  
\begin{lstlisting}[ language=caml]

(* Bounded verification of Lemma 4.1 R2 for dimension 8 *)

lemma extreme_value_lemma_r2_len_8 x1 x2 x3 x4 x5 x6 x7 x8
    y1 y2 y3 y4 y5 y6 y7 y8
    a1 a2 a3 a4 a5 a6 a7 a8 =
    let w_x, w_y, a = [x1; x2; x3; x4; x5; x6; x7; x8],
    [y1; y2; y3; y4; y5; y6; y7; y8],
    [a1; a2; a3; a4; a5; a6; a7; a8]
    in
    extreme_value_precondition_r2 w_x w_y a
    ==> extreme_value_postcondition w_x w_y a
    [@@unroll 200]

  \end{lstlisting}

Imandra informs us that the unrolling tactic managed to produce the full proof.
Moreover, if there was a counterexample up to this bound, Imandra would present
it to us (and reflect it in the runtime). In fact, this is precisely how we
strengthened some of the conditions for our proofs.

  \begin{lstlisting}[ language=caml]

# lemma extreme_value_lemma_r2_len_2 x1 x2 y1 y2 a1 a2 =
    let w_x, w_y, a = [x1; x2],
                      [y1; y2],
                      [a1; a2]
    in
    extreme_value_precondition_r2 w_x w_y a
    ==> extreme_value_postcondition w_x w_y a
  ;;
val extreme_value_lemma_r2_len_2 :
  real -> real -> real -> real -> real -> real -> bool = <fun>
Theorem proved.
\end{lstlisting}



\section{Algebraic Data Types and Records for Matrices}\label{ap:records}

\subsection{Algebraic Data Types for  Real-Valued Matrices}

The first alternative is to introduce an algebraic data type that allows the matrix functions to return either reals or integers. 

\begin{lstlisting}[language=caml]
type arg =
	| Rows
	| Cols
	| Value of int * int
	| Default

type 'a res = 
	| Int of int
	| Val of 'a

type 'a t = arg -> 'a res
\end{lstlisting}

This allows a form of polymorphism, but it also introduces pattern matching each time we query a value from the matrix. For instance, in order to use dimensions as indices to access a matrix element we have to implement the following \lstinline{nth_res} function:  

\begin{lstlisting} [language=caml]
let nth_res (m: 'a t) (i: 'b res) (j: 'c res): 'a res = match (i, j) with 
	| (Int i', Int j') -> m (Value (i', j'))
	| _                -> m Default
\end{lstlisting}

The simplicity and efficiency of the functional implementation is lost, and some indicative runs on the ACAS Xu challenge show reduction in performance.

\subsection{Records}

Standard OCaml records are available in Imandra, though they do not support functions as fields. This is because all records are data values which must support a computable equality relation, and in general one cannot compute equality on functions. Internally in the logic, records correspond to algebraic data types with a single constructor and the record fields to named constructor arguments. Like product types, records allow us to group together values of different types, but with convenient accessors and update syntax based on field names, rather than position. This offers the possibility of polymorphism for our matrix type. 

The approach here is similar to the one in Section~\ref{sec:acas}: matrices are stored as mappings between indices and values, which allows for constant-time access to the elements. However, instead of having the mapping be implemented as a function, here we implement it as a \lstinline{Map}, i.e. an unordered collection of (key;value) pairs where each key is unique, so that this ``payload'' can be included as the field of a record. 

\begin{lstlisting}[language=caml]
type 'a t = {
	rows: int;
	cols: int;
	vals: ((int*int), 'a) Map.t;
}
\end{lstlisting}

We can then use a convenient syntax to create a record of this type. For instance, a weights matrix from one of the ACAS Xu networks can be implemented as:

\begin{lstlisting}[language=caml]
let layer6_map =
	Map.add (0,10) (0.05374) @@
	Map.add (0,20) (0.05675) @@
	...
	Map.const 0.

let layer6_matrix = {
	rows = 5;
	cols = 51;
	vals = layer6_map;
}
\end{lstlisting}

Note that the matrix dimensions (and the underlying map's keys) are indeed encoded as integers, whereas the weights' values are reals. 

Similarly to the previous implementations, we define a number of useful matrix operations which will be used to define general neural network layer functions. For instance, the \lstinline{map2} function is defined thus:

\begin{lstlisting}[language=caml, label={lst:map2_records}]
let rec map2_rec (m: 'a t) (m': 'b t) (f: 'a -> 'b -> 'c) (cols: int) (i: int) (j: int) (res: ((int*int), 'c) Map.t): ((int*int), 'c) Map.t =
		let dec i j = 
			if j <= 0 then (i-1, cols) else (i,j-1)     
		in
		if i <= 0 && j <= 0 then (
			res
		) else (
			let (i',j') = dec i j in
			let new_value = f (nth m (i',j')) (nth m' (i', j')) in
			let res' = Map.add' res (i',j') new_value in
			map2_rec m m' f cols i' j' res'
		)
[@@adm i,j]

let map2 (f: 'a -> 'b -> 'c) (m: 'a t) (m': 'b t) : 'c t = 
	let rows = max (m.rows) (m'.rows) in
	let cols = max (m.cols) (m'.cols) in
	let vals = map2_rec m m' f cols rows cols (Map.const 0.) in
	{
		rows = rows;
		cols = cols;
		vals = vals;
	}
\end{lstlisting} 

Compared to the list implementation, this implementation has the benefit of providing constant-time access to matrix elements. However, compared to the implementation of matrices as functions, it uses recursion to iterate over matrix values which results in a high number of case-splits. This in turn results in lower scalability.
Compared to the previous section's results, none of the verification tests on pruned ACAS Xu benchmarks that terminated with the functional matrix implementation terminated with the records implementation.

Moreover, we can see in the above function definition that we lose considerable conciseness and readability.

In the end, the main interest of this implementation is its offering polymorphism. In all other regards, the functional implementation seems preferable.

\end{document}